\documentclass[]{aa}
\usepackage[varg]{txfonts}
\usepackage{graphicx}
\usepackage{amsmath}
\bibpunct{(}{)}{;}{a}{}{,} 

\begin{document}

\title{The central dynamics of \object{M3}, \object{M13}, and \object{M92}: Stringent limits on the masses of intermediate-mass black holes\thanks{Based on observations collected at the Centro Astron\'{o}mico Hispano Alem\'{a}n (CAHA) at Calar Alto, operated jointly by the Max-Planck Institut f\"ur Astronomie and the Instituto de Astrof\'{\i}sica de Andaluc\'{\i}a (CSIC)}\thanks{Tables D1 to D6 are only available in electronic form at the CDS via anonymous ftp to cdsarc.u-strasbg.fr (130.79.128.5) or via http://cdsweb.u-strasbg.fr/cgi-bin/qcat?J/A+A/}}

\author{S.~Kamann\inst{1}\fnmsep\inst{2}
  \and L.~Wisotzki\inst{1}
  \and M.~M.~Roth\inst{1}
  \and J.~Gerssen\inst{1}
  \and T.-O.~Husser\inst{2}
  \and C.~Sandin\inst{1}
  \and P.~Weilbacher\inst{1}
}
\institute{Leibniz-Institut f\"ur Astrophysik Potsdam (AIP), An der Sternwarte 16, 14482 Potsdam, Germany
\and{Institut f\"ur Astrophysik, Universit\"at G\"ottingen, Friedrich-Hund-Platz 1, 37077 G\"ottingen, Germany\\ \email{skamann@astro.physik.uni-goettingen.de}}
}

\date{Received / Accepted}

\abstract{
We used the PMAS integral field spectrograph to obtain large sets of radial velocities in the central regions of three northern Galactic globular clusters: \object{M3}, \object{M13}, and \object{M92}. By applying the novel technique of crowded field 3D spectroscopy, we measured radial velocities for about $80$ stars within the central $\sim10\arcsec$ of each cluster. These are by far the largest spectroscopic datasets obtained in the innermost parts of these clusters up to now. To obtain kinematical data across the whole extent of the clusters, we complement our data with measurements available in the literature.
We combine our velocity measurements with surface brightness profiles to analyse the internal dynamics of each cluster using spherical Jeans models, and investigate whether our data provide evidence for an intermediate-mass black hole in any of the clusters.
The surface brightness profiles reveal that all three clusters are consistent with a core profile, although shallow cusps cannot be excluded. We find that spherical Jeans models with a constant mass-to-light ratio provide a good overall representation of the kinematical data. A massive black hole is required in none of the three clusters to explain the observed kinematics. Our $1\sigma$ ($3\sigma$) upper limits are $5\,300\ \mathrm{M}_\odot$ ($12\,000\ \mathrm{M}_\odot$) for \object{M3}, $8\,600\ \mathrm{M_\odot}$ ($13\,000\ \mathrm{M}_\odot$) for \object{M13}, and $980\ \mathrm{M}_\odot$ ($2\,700\ \mathrm{M}_\odot$) for \object{M92}. A puzzling circumstance is the existence of several potential high velocity stars in \object{M3} and \object{M13}, as their presence can account for the majority of the discrepancies that we find in our mass limits compared to \object{M92}.}

\titlerunning{The central dynamics of \object{M3}, \object{M13} and \object{M92}}
\authorrunning{S.~Kamann et al.}

\defcitealias{2013A&A...549A..71K}{Paper~I}
\newcommand{\Otwo}{\mbox{O$_\mathrm{2}$\,}}

\maketitle

\section{Introduction}
\label{sec:introduction}
Over the last years, the search for intermediate-mass black holes (IMBHs) has attracted remarkable attention. With $\sim10^2\text{--}10^4$ solar masses, they would bridge the gap from stellar-mass black holes to supermassive ones (SMBHs). Constraining their population statistics might answer the question how SMBHs assemble their masses. The scaling relations observed between SMBHs and fundamental properties of their host bulges, such as luminosities \citep{1995ARA&A..33..581K}, stellar masses \citep{1998AJ....115.2285M,2003ApJ...589L..21M,2004ApJ...604L..89H} or stellar velocity dispersions \citep{2000ApJ...539L...9F,2000ApJ...539L..13G,2009ApJ...698..198G} suggest that their growth is closely linked to the evolution of the host galaxy \citep[][but see also \citealt{2011ApJ...734...92J}]{1998A&A...331L...1S}. The progenitors of SMBHs are likely to be found in the building blocks of present-day galaxies. The close connection of Galactic globular clusters to the build-up of the Milky Way was proposed already by \citet{1978ApJ...225..357S} and is supported by cosmological simulations \citep[e.g.,][]{2006MNRAS.368..563M}. Globular clusters can therefore be considered as promising candidates to host IMBHs.
Interestingly, a straightforward extrapolation of the SMBH scaling relations to the properties of globular clusters also yields black hole masses in the range $10^2\text{--}10^5$ solar masses. Clearly, such an extrapolation is a huge simplification, and there may be evidence that the common relations disagree with the observations already in the regime of low-mass galaxies \citep{2010ApJ...721...26G}.

The runaway merging of massive stars in the early phases of cluster evolution has been suggested as a formation channel for IMBHs in dense star clusters \citep{2002ApJ...576..899P}. However, it has been argued \citep[e.g.][]{2009A&A...497..255G} that strong stellar winds restrict the resulting black holes to stellar masses ($\gtrsim10~\mathrm{M_\odot}$). An alternative formation scenario involves the collapse of massive population III stars \citep[e.g.][]{2001ApJ...551L..27M}.

Observational evidence for (but also against) the existence of IMBHs is still extremely scant. Gas accretion onto such a black hole would allow for its detection using radio or X-ray observations, such as in the case of the IMBH candidate HLX-1 in the galaxy \object{ESO~243-49} \citep{2009Natur.460...73F}. \object{G1} in \object{M31}, the most massive known globular cluster in the local group, was detected in both radio \citep{2007ApJ...661L.151U} and X-ray observations \citep{2004ApJ...616..821T,2010MNRAS.407L..84K}, although \citet{2012ApJ...755L...1M} could not confirm the radio detection. Recent observations by \citet{2012ApJ...750L..27S} place stringent upper limits on the amount of radio emission coming from Galactic globular clusters. Translating those into mass limits, however, requires making assumptions about the accretion physics that are not well understood.

Several authors investigated the possibility that IMBHs imprint their presence onto photometrically observable properties of a globular cluster. \citet{2005ApJ...620..238B} and \citet{2011ApJ...743...52N} found that a massive black hole should produce a shallow cusp in the central surface brightness profile of the surrounding cluster. A large ratio of core to half-mass radius was suggested as indirect evidence for the presence of an IMBH by \citet{2007MNRAS.374..857T}. Furthermore, the existence of extreme horizontal branch stars has been proposed as a tracer for IMBHs by \citet{2007MNRAS.381..103M}. \citet{2008ApJ...686..303G} investigated the influence of black holes on mass segregation among the cluster stars.

Arguably, the most direct way to not only find massive black holes but also obtain their masses is the detection of their kinematic fingerprints. The best SMBH mass estimates were obtained from stellar kinematics in our own galaxy \citep{2009ApJ...692.1075G} or gas kinematics in \object{NGC~4258} \citep{2005ApJ...629..719H}. However, obtaining meaningful kinematic measurements in the central regions of globular clusters is a challenging task: While the measurement of individual stellar velocities is hampered by crowding, integrated-light analyses can be significantly affected by shot noise from the few brightest giants \citep{1997A&A...324..505D}. Kinematic studies of the centres of globular clusters remained ambiguous so far, in some cases even contradictory conclusions were reached for the same clusters. A prominent example is \object{$\omega$~Centauri}, for which \citet{2010ApJ...710.1032A} and \citet{2010ApJ...710.1063V} find no evidence for an IMBH while \citet{2010ApJ...719L..60N} and \citet{2012A&A...538A..19J} claim the detection of one, with a mass of $4\times10^4\ \mathrm{M_\odot}$. Similarly, the detection of an IMBH with $17\,000$ solar masses in \object{NGC~6388} by \citet{2011A&A...533A..36L} was not confirmed by \citet{2013ApJ...769..107L}, who obtain an upper limit of $2\,000\ \mathrm{M_\odot}$ instead. Further detections of IMBHs have been reported in a small number of massive clusters, \object{G1} \citep{2005ApJ...634.1093G} among them. In a sample recently studied by \citet{2012A&A...542A.129L,2013A&A...552A..49L}, the kinematics in $2$ out of $7$ clusters suggested the presence of an IMBH, while for the remaining clusters upper limits of typically $>1\,000$ solar masses were derived. Similar mass limits were also reported in other studies, carried out for \object{M15} by \citet{2002AJ....124.3270G} and \citet{2006ApJ...641..852V}, for \object{47~Tuc} by \citet{2006ApJS..166..249M} or for \object{NGC~6266} by \citet{2012ApJ...745..175M}.

Clearly, a conclusive picture of which globular clusters host IMBHs is not established yet. Consequently, the question whether the scaling relations established for SMBHs can be extrapolated into the regime of globular clusters is also unanswered. So far the results are still consistent with it. Alternatively, IMBHs might follow different scaling relations, as suggested by \citet{2007MNRAS.381..103M}. More observations are the only way to make progress here.

So far, all claimed IMBH detections come from integrated light spectroscopy, whereas studies based on the kinematics of resolved stars derived upper limits which in some cases are in conflict with the detections from the former approach. This may suggest that the influence of the few brightest stars still hampers the integrated light measurements. In any case, resolving this situation requires new techniques for spectroscopy in crowded stellar fields and a better understanding of its capabilities and limitations. In \citet[][hereafter \citetalias{2013A&A...549A..71K}]{2013A&A...549A..71K}, we recently presented a new method for analysing integral field spectroscopy (IFS) data of such fields. It extends the established analysis techniques for crowded field photometry into the domain of three-dimensional datacubes by fitting a wavelength dependent point spread function (PSF) to deblend stellar spectra. In this paper, we apply this technique to IFS data of three Galactic globular clusters, \object{M3}, \object{M13}, and \object{M92}, with the aim  of constraining the presence of intermediate-mass black holes in the objects. The paper is organized as follows. After laying out the the target selection, the observations, and their reduction in Sects.~\ref{sec:sample} to \ref{sec:reduction}, we describe the analysis of the photometric (Sect.~\ref{sec:photometry}) and the kinematic data (Sect.~\ref{sec:kinematic}). In Sect.~\ref{sec:dynamics}, the cluster dynamics are analysed and the search for intermediate-mass black holes in the clusters via Jeans modelling is performed. Our findings are discussed in Sect.~\ref{sec:discussion}, and we provide our conclusions in Sect.~\ref{sec:conclusions}.

\section{Target selection}
\label{sec:sample}

\begin{figure}
 \resizebox{\hsize}{!}{\includegraphics{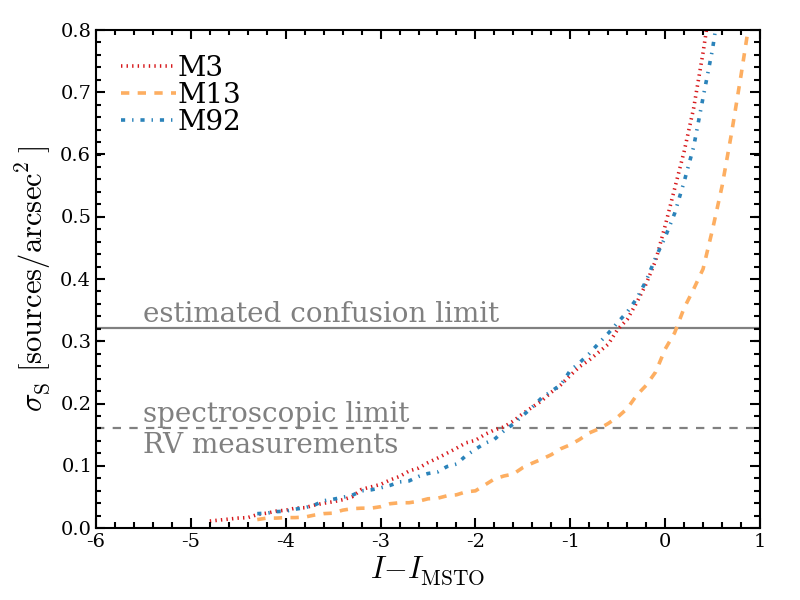}}
 \caption{Central surface number densities of sources brighter than a given magnitude for the three clusters of our sample, colour-coded as indicated in the upper left corner of the plot. Magnitudes are given relative to the main sequence turn-off (MSTO) in each cluster. The solid horizontal line corresponds to the source density where we expect to hit the confusion limit. Below the density indicated by a dashed line the deblended spectra will likely have a sufficient signal-to-noise ratio for radial velocity determination.}
 \label{fig:sample:mag_vs_density}
\end{figure}

\begin{table*}
 \caption{Fundamental properties of the observed clusters}
 \label{tab:pmas_clusters}
 \centering
 \begin{tabular}{ c c c c c c c c }
  \hline\hline
  Cluster & RA & Dec & distance & $M_{V,\mathrm{t}}$ & $v_\mathrm{sys}$ & $\sigma_\mathrm{0} $ & [Fe/H] \\
  & J2000 & J2000 & kpc & mag & $\mathrm{km\,s^{-1}}$ & $\mathrm{km\,s^{-1}}$ & \\ \hline
  \object{M3} (NGC~5272) & 13 42 11.62 & +28 22 38.2 & 10.2 & -8.88 & -147.6 & 5.5 & -1.50 \\
  \object{M13} (NGC~6205) & 16 41 41.24 & +36 27 35.5 & 7.1 & -8.55 & -244.2 & 7.1 & -1.53 \\
  \object{M92} (NGC~6341) & 17 17 07.39 & +43 08 09.4 & 8.3 & -8.21 & -120.0 & 6.0 & -2.31 \\ \hline
 \end{tabular}
 \tablefoot{The values are taken from the database of \citet{1996AJ....112.1487H}.}
\end{table*}

We searched the catalogue of \citet{1996AJ....112.1487H} for globular clusters visible from Calar Alto observatory and selected the objects that were best suited for an analysis. This was done based on the following criteria. As we showed in \citetalias{2013A&A...549A..71K}, the confusion limit below which individual stellar spectra cannot be resolved any more is about $0.4$ stars per resolution element. Assuming a typical seeing of $1.0\arcsec$, this implies that we can resolve stars down to a density of $0.3\,\mathrm{arcsec}^{-2}$. Using the $V$- and $I$-band photometry obtained in the ACS Survey of Galactic globular clusters (\citealt{2007AJ....133.1658S}; \citealt{2008AJ....135.2055A}), we determined the $I$-band magnitudes where the confusion limit is reached by counting the stars in the central 20$\arcsec$ of each cluster. Only clusters where spectroscopic observations down to this limit are feasible with a 3.5m telescope were considered. Additionally, we restricted ourselves to clusters with a confusion limit in the central region close to the main sequence turn-off (MSTO). The motivation for this selection criterion was that the number of stars per magnitude bin increases strongly below this point and we planned to analyse not only the resolved stars but also the unresolved stellar component after we subtracted the bright stars. Now, if we can resolve nearly all stars on the red giant branch of the cluster, we ensure that within the unresolved stellar component many stars have similar brightnesses and that the integrated light of that component is not dominated by few bright stars. To illustrate this second criterion, we show in Fig.~\ref{fig:sample:mag_vs_density} the source density as a function of $I$-band magnitude for the three clusters that best fulfilled our selection criteria, \object{M3} (\object{NGC~5272}), \object{M13} (\object{NGC~6205}) and \object{M92} (\object{NGC~6341}). The expected confusion limit in each cluster is indicated. We also provide the ``usefulness'' limit, motivated by the fact found in \citetalias{2013A&A...549A..71K} that roughly half of the deblended spectra allow for a measurement of the radial velocity of the star. The analysis of the unresolved stellar component will be presented in a later publication and is not discussed in this study. Table \ref{tab:pmas_clusters} summarizes some fundamental properties of our sample clusters. The three targets are among the most massive clusters visible in the northern sky and are also located at relatively small distances from the sun.

We did not perform a preselection based on photometric properties, such as the central slopes of the surface brightness profiles of the clusters. \citet{2005ApJ...620..238B} suggested that the central slopes might be used as an indicator for the presence of intermediate-mass black holes in the sense that clusters harbouring one should have a shallow cusp in their surface brightness profiles. However, we wanted to avoid biases for selecting ``favourable'' objects.

Dynamical studies have been previously performed for each of the three clusters. The kinematics in \object{M3} were investigated already by \citet{1979AJ.....84..752G} who found that King-Mitchie models provided a valid framework for the kinematical data available back then. The authors suggested a transition from an isotropic core to an anisotropic outer region occurring at $\sim15$ core radii ($\sim5\arcmin$). This finding was confirmed by \citet{1979AJ.....84.1312C} from proper motions. In a similar fashion to \citet{1979AJ.....84..752G}, \citet{1987AJ.....93.1114L} studied the cluster \object{M13} and found a transition to anisotropy occurring at $\sim5$ core radii. Furthermore, the collected radial velocity data showed evidence for rotation at larger radii. \citet{1992AJ....104.2104L} combined the kinematical data of \citet{1987AJ.....93.1114L} with proper motions collected by \citet{1979AJ.....84..774C} to infer the mass and stellar content of \object{M13}, by analytically solving the spherical Jeans equation under the assumption of a Plummer density profile.

More recently, \citet{2005ApJS..161..304M} performed dynamical modelling on a large sample of 57 clusters, including \object{M3}, \object{M13} and \object{M92}. To infer structural parameters, three different models were used that mainly varied in their behaviour at large radii, thus corresponded to different strengths of a tidal cut-off. For the three targets investigated in this study, models with a moderate truncation (Wilson models) turned out to be a valid choice and we will refer to those when comparing our results with those of \citeauthor{2005ApJS..161..304M}. \object{M92} was also included in the sample studied by \citet{2012A&A...539A..65Z} who found it to be a relaxed cluster well described by an isotropic King model.

We note that \object{M13} was identified by \citet{2007MNRAS.381..103M} as a possible candidate to host an IMBH according to its central surface brightness profile and extended horizontal branch.

\section{Observations}
\label{sec:observations}

\begin{table*}
 \caption{Summary of the PMAS observations}
 \label{tab:pmas_observations}
 \centering
 \begin{tabular}{c r c r c c c }
 \hline\hline
 Run ID & \multicolumn{3}{c}{Observing dates} & \multicolumn{2}{c}{HJD} & Targets \\
 \hline
 173 & 14 Mar 2010 & -- & 18 Mar 2010 & 2455273 & 2455275 & \object{M3} \\
 184 & 07 May 2011 & -- & 09 May 2011 & 2455690 & 2455691 & \object{M13} \\
 191 & 07 Jul 2011 & -- & 10 Jul 2011 & 2455750 & 2455753 & \object{M13}, \object{M92} \\
 197 & 01 Oct 2011 & -- & 02 Oct 2011 & 2455836 & 2455838 & \object{M92} \\
 \hline
 \end{tabular}
\end{table*}

We observed the three clusters using the PMAS instrument \citep{2005PASP..117..620R} mounted on the Calar Alto 3.5m telescope. PMAS is an optical integral field spectrograph without adaptive optics. The data were obtained in 4 different observing runs in 2010 and 2011. In Table~\ref{tab:pmas_observations} we give a summary of the different observing runs. All observations were carried out using a spatial pixel (``spaxel'') scale of $0\farcs5$ on the sky and the R1200 grating in first order mounted backwards. With this configuration, we covered the wavelength range from $7\,300\,\text{\AA}$ to $8\,900\,\text{\AA}$ and achieved a spectral resolution of $R=\lambda/\Delta\lambda\sim7\,000$ around the infrared \ion{Ca}{ii}-triplet. The precise value of $R$ varied with wavelength and fibre. For \object{M13} and \object{M92} some observations were repeated in a later observing run, which enabled us to search for stars with variable radial velocity.

PMAS covers a contiguous area of $8\arcsec\times8\arcsec$ in the sky in the selected mode. For each of the observed targets, the aim was to completely cover the central region of the cluster, out to a radius of about $15\arcsec$. Due to weather losses, however, some fields could not be observed and the achieved coverage varied from cluster to cluster. Additionally, the final acquisition of the PMAS IFS had to be performed manually. In a crowded stellar field like a globular cluster, this procedure has an error of at least $1\arcsec-2\arcsec$ which constrained our final pointing accuracy.

Except for the first run (ID 173), science observations were alternated with shorter observations of blank sky fields to allow for a better subtraction of sky lines during the data reduction.

\section{Data reduction}
\label{sec:reduction}

The bulk part of the data reduction was performed using \textsc{p3d}\footnote{\url{http://p3d.sourceforge.net/}}, a dedicated software package to reduce fibre-fed integral field observations \citep{2010A&A...515A..35S}. We carried out the basic parts of the reduction cascade, such as bias subtraction, tracing, and extraction of the spectra, wavelength calibration, and rebinning the data onto a regular grid. Recently, the dedicated cosmic-ray-rejection routine \textsc{pyCosmic} \citep{2012A&A...545A.137H} has been implemented into \textsc{p3d} and we followed the suggestions by \citeauthor{2012A&A...545A.137H} to detect cosmic ray hits in the raw data.

We used an optimal extraction method to obtain the individual spectra. Although optimal extraction is not strictly necessary for data observed with the PMAS lens array (the gaps between individual fibres are sufficiently large that crosstalk is a negligible issue), it has the big advantage over the simpler boxcar extraction that pixels affected by cosmic rays can be masked out, so we did not rely on any interpolation scheme to assign values to those pixels.

All of our data were observed using the new PMAS CCD \citep{2010SPIE.7742E...7R}, which exhibits some features that required special treatment during the reduction. Some steps of the data reduction were therefore performed outside of \textsc{p3d} and are outlined in Appendix \ref{app:reduction}.

\section{Photometry}
\label{sec:photometry}

\subsection{Generating a complete source list}

Our analysis depends crucially on the availability of high quality photometric data, in particular near the cluster centres, for two reasons. First, our analysis of the IFS data relies on a reference catalogue containing locations and magnitudes of the stars. Second, the Jeans modelling approach used in Sect.~\ref{sec:dynamics} below uses the surface brightness profile of the clusters as input.

To obtain a high quality reference catalogue around the centres of the clusters, we started again from the photometry obtained in the ACS Survey of Galactic globular clusters, certainly the most comprehensive photometric dataset available for our clusters. In Fig.~\ref{fig:appb:corrected_cmds} we show a colour magnitude diagram of the three clusters based on this data. One common issue of ACS observations is that the brightest stars are often heavily saturated and cause strong bleeding artefacts. Although the ACS survey included an observing strategy to overcome this effect as much as possible and saturated stars were treated separately in the analysis by \citet{2008AJ....135.2055A}, we decided to cross-check with a second dataset. In fact, a single PMAS datacube only contains 256 spaxels, so even a single star that is significantly detectable yet missing in the reference catalogue would have a strong effect on the analysis. We performed a search and correction for such stars using WFPC2 data as outlined in Appendix~\ref{app:catalog_correction}.

\subsection{Determination of the cluster centres}
\label{sec:photometry:centre}

For the later analysis, a proper determination of the central coordinates of each cluster was important. Measured cluster properties such as the surface brightness profile or the velocity dispersion profile can deviate significantly from the intrinsic ones if the true centre is offset from the assumed one. Several methods were proposed in the literature to determine the (photometric) centre of a globular cluster, such as fitting ellipses to the number density contours of the cluster and determining the centre by averaging the centres of the different ellipses. Other methods exploit the symmetry of the density around putative centres. For the clusters in the ACS survey, these methods were applied by \citet{2010AJ....140.1830G}. Since our photometry is based on more or less the same data (the few missing stars have marginal influence) we adopted the centres reported by \citeauthor{2010AJ....140.1830G}. They are also provided in Table~\ref{tab:pmas_clusters}.

The central coordinates deviate slightly from those measured by \citet{2006AJ....132..447N}. The offsets are $3\farcs9$ for \object{M3}, $2\farcs4$ for \object{M13}, and $0\farcs9$ for \object{M92}. A reason for this might be that all targets have rather large cores \citep[$>\!10\arcsec$,][]{1996AJ....112.1487H} so that measurements of the centres are not well constrained. The measurements performed by \citeauthor{2006AJ....132..447N} were based on WFPC2 data that cover a smaller fraction of the central region than ACS does. Therefore, we relied on the \citeauthor{2010AJ....140.1830G} centres.

\subsection{Surface brightness profiles}
\label{sec:photometry:sbprofiles}

\begin{figure*}
 \centering\includegraphics[width=17cm]{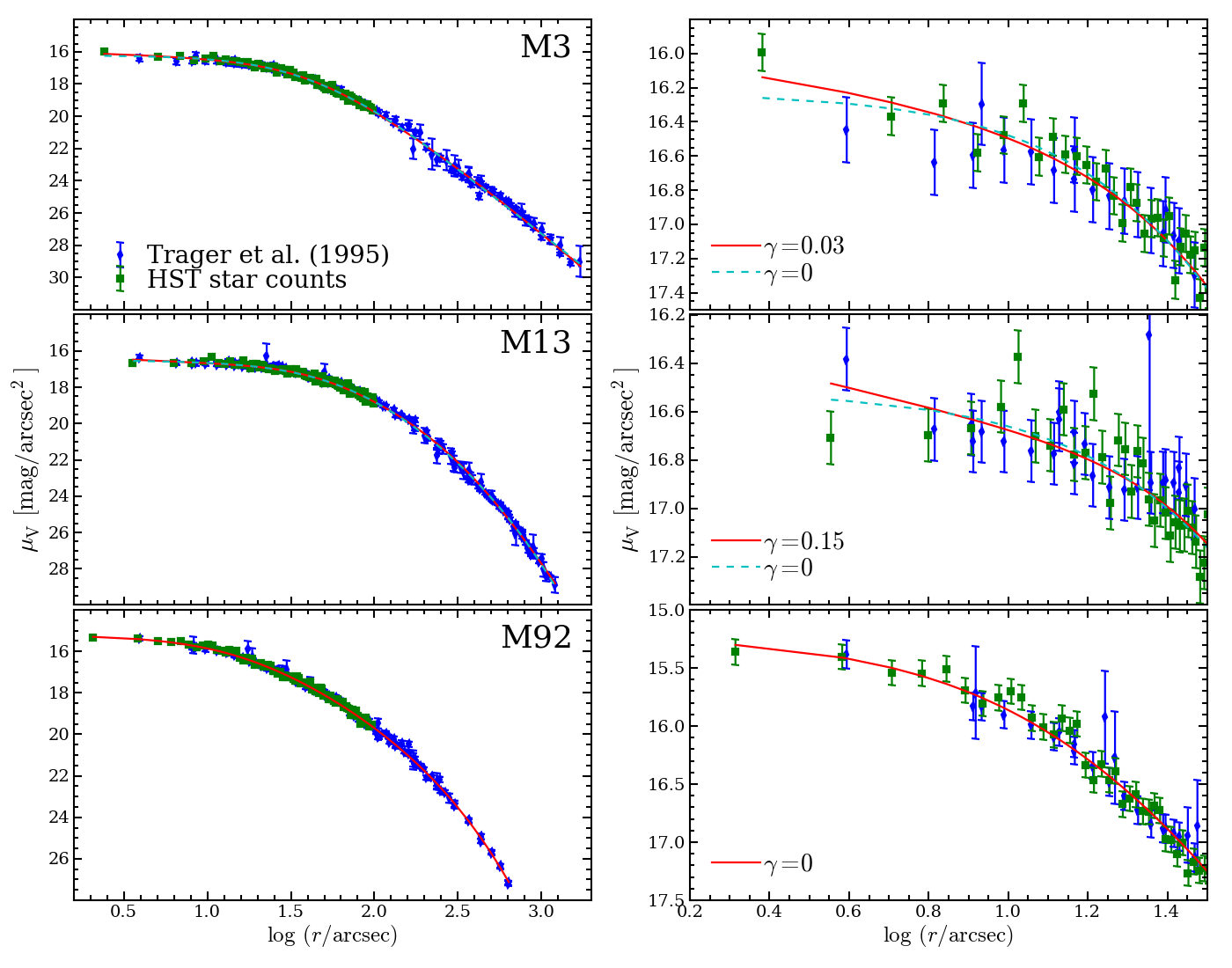}
 \caption{The surface brightness profiles of the clusters \object{M3}, \object{M13}, and \object{M92} are presented over the entire extent of each cluster on the left hand side. The right hand side zooms into the central region of each cluster. Measurements based on star counts in the HST data are shown using green squares while blue diamonds represent the data of \citet{1995AJ....109..218T}. A solid red line shows the best-fit extended Nuker profile. For \object{M3} and \object{M13}, we also show a Nuker profile with a core (i.e. $\gamma=0$) as a cyan dashed line. The surface brightness profiles have been corrected for extinction.}
 \label{fig:photometry:sbprofiles}
\end{figure*}

To generate surface brightness profiles from the photometric catalogues, we proceeded as follows. We counted the stars in concentric annuli around the cluster centre. The size of each annulus was chosen so that it contained 100 stars brighter than a magnitude $V=19$ in order to have the same number statistics in each bin. Note that except for the cut in magnitude, each star was weighted equally, so this method is robust against shot noise from a few bright giant stars. To translate the measured projected number density into a surface brightness, we assumed a constant conversion factor with radius. By doing so we neglected any mass segregation among our sample stars. However, as the stars that we probe fall in a narrow range in mass, mass segregation is not a major concern. 

Our approach allowed us to constrain the central profile for approximately the same subsets of stars for which we expected to obtain kinematic measurements. The adapted magnitude cut of $V=19$ is slightly fainter than the assumed limit of our spectroscopic observations but this should have no effect. When we used a brighter magnitude cut the shape of the profile did not change significantly, only the number statistics got poorer. This also shows that the catalogue is quite complete at this brightness levels. 

Our catalogues only cover the central regions of the clusters. To obtain profiles across the whole extent of each object, we complemented them with the data of \citet{1995AJ....109..218T}. No uncertainties are provided with the data, only a relative weight is assigned to each datapoint. \citet{2005ApJS..161..304M} proposed a way to obtain meaningful uncertainties by scaling the weights with a constant factor until the measurements are consistent with a smooth function. We followed that approach and used the same scaling factors as \citeauthor{2005ApJS..161..304M} and also omitted datapoints that were assigned a weight $<0.15$. To merge the two catalogues, the number density profiles from the HST data were shifted vertically to match the surface brightness profiles of \citeauthor{1995AJ....109..218T} in the region where the two overlap.

The resulting profiles were corrected for interstellar extinction and fitted with an extended Nuker profile, the functional form of which is given by
\begin{equation}
 \label{eq:ext_nuker}
 I(r) = I_\mathrm{b}\,2^{\left(\frac{\beta-\gamma}{\alpha}\right)}\,\left(\frac{r}{r_\mathrm{b}}\right)^{-\gamma}\,\left[1 + \left(\frac{r}{r_\mathrm{b}}\right)^\alpha\right]^{-\left(\frac{\beta-\gamma}{\alpha}\right)}\,\left[1 + \left(\frac{r}{r_\mathrm{c}}\right)^{\delta}\right]^{-\left(\frac{\epsilon-\beta}{\delta}\right)}\,.
\end{equation}
The ``original'' Nuker profile (omitting the second term in square brackets) was initially proposed by \citet{1995AJ....110.2622L} to describe the surface brightness profiles of galaxies that showed a power-law cusp $I \propto r^{-\gamma}$ towards the centre and a logarithmic decline $\propto\beta$ for radii >$r_\mathrm{b}$. The generalized version given in Eq.~\ref{eq:ext_nuker} was used by \citet{2010ApJ...710.1063V} to describe the surface brightness profile of \object{$\omega$~Centauri}. It has an additional break radius $r_\mathrm{c}$ where the logarithmic slope changes from $\beta$ to $\epsilon$. The physical justification for this second break is that many clusters have rather well-defined truncation radii.

In Fig.~\ref{fig:photometry:sbprofiles} we show the results from fitting Eq.~\ref{eq:ext_nuker} to the surface brightness profiles. The parameters of each best-fit Nuker profile are given in Table~\ref{tab:photometry:nuker_fits}. The fits provide a good global representation of the measured profiles. In all three cases we obtain $\chi^2$ values indicating that the fitted profile is fully consistent with the data (cf. Table~\ref{tab:photometry:nuker_fits}).

\begin{table*}
 \caption{Parameters of the best-fit Nuker profiles}
 \label{tab:photometry:nuker_fits} 
 \centering
 \begin{tabular}{c c c c c c c c c c c c c }
\hline\hline
ID & $\mu_\mathrm{b}$ & $r_\mathrm{b}$ & $r_\mathrm{c}$ & $\alpha$ & $\beta$ & $\gamma$ & $\delta$ & $\epsilon$ & $\chi^2$ & $N_\mathrm{dof}$ & $m_\mathrm{V, data}$ & $m_\mathrm{V, Nuker}$ \\ 
   & mag/arcsec$^2$  & arcsec & arcsec &  &  &  &  &  &  & \\
\hline
M3  & 16.8 & 41.5 &  273.5 & 2.10 & 2.15 & 0.03 & 0.57 & 3.94 & 167.7 & 161 & 6.50 & 6.55 \\
M13 & 17.2 & 39.8 &  353.2 & 2.72 & 1.39 & 0.15 & 1.46 & 6.06 & 242.4 & 245 & 5.88 & 5.89 \\
M92 & 16.1 & 15.1 & 2979.1 & 1.47 & 1.69 & 0.00 & 0.95 & 22.8 & 183.5 & 172 & 6.46 & 6.49 \\

\hline
\end{tabular}
\end{table*}

An important parameter regarding the presence of intermediate-mass black holes is the central slope of the profile, $\gamma$. Our fits result in low $\gamma$ values, in fact \object{M92} is best fit by a core (i.e. $\gamma=0$) profile, and for the other two clusters, the deviation from core profiles is only mild (cf. Fig.~\ref{fig:photometry:sbprofiles}, right). However, as pointed out by \citet{2010ApJ...710.1063V}, $\gamma$ is not very well constrained in a fit of the whole surface brightness profile because its influence is limited to the innermost few datapoints. To obtain realistic confidence intervals on $\gamma$, we fitted the data within a distance of $0.5r_\mathrm{b}$ to the centre with a simple power-law and obtained the range of slopes consistent with the data. We only allowed for $\gamma$ values that still provided a reasonable fit to the overall profile within 90\% confidence. This calculation yields $\gamma<0.54$ in \object{M3}, $\gamma<0.34$ in \object{M13} and $\gamma<0.43$ in \object{M92}.

In summary, none of our clusters shows significant evidence for a central cusp in its surface brightness profile. This result is in agreement with other studies. The central slopes reported by \citet{2006AJ....132..447N} for the three targets are smaller than their respective uncertainties. More recently, \citet{2013ApJ...774..151M} found that King and Wilson profiles provide an adequate representation of the clusters and detected no evidence of any central cusp.

A comparison of the core radii obtained by us with those listed in the catalogue of \citet[][M3: $22\farcs2$, \object{M13}: $37\farcs2$, \object{M92}: $15\farcs6$]{1996AJ....112.1487H} shows remarkable agreement for \object{M13} and \object{M92}, only the value we get for \object{M3} is about twice as large. Our measurements correspond to physical sizes of $2.1\ \mathrm{pc}$ (\object{M3}), $1.4\ \mathrm{pc}$ (\object{M13}), and $0.6\ \mathrm{pc}$ (\object{M92}).

We performed another consistency check by calculating the integrated magnitude of each target, either by numerical integration of the data or Eq.~\ref{eq:ext_nuker}. Again, the different values, included in Table~\ref{tab:photometry:nuker_fits}, are largely consistent with one another.

The combination of a large core radius and a shallow surface brightness cusp led \citet{2007MNRAS.381..103M} to the conclusion that \object{M13} is a good candidate to host an IMBH. Our photometric analysis confirms the large core. Also, the confidence interval for the central slope that we obtain is consistent with the value $\gamma\lesssim0.25$ suggested by \citeauthor{2007MNRAS.381..103M} and others \citep{2005ApJ...620..238B,2011ApJ...743...52N}. Our analysis further suggests that the photometric properties of \object{M3} are not so different from those of \object{M13}, in particular with respect to the large core radius.

\section{Kinematic data}
\label{sec:kinematic}

\begin{figure*}
 \centering
  \includegraphics[width=17cm]{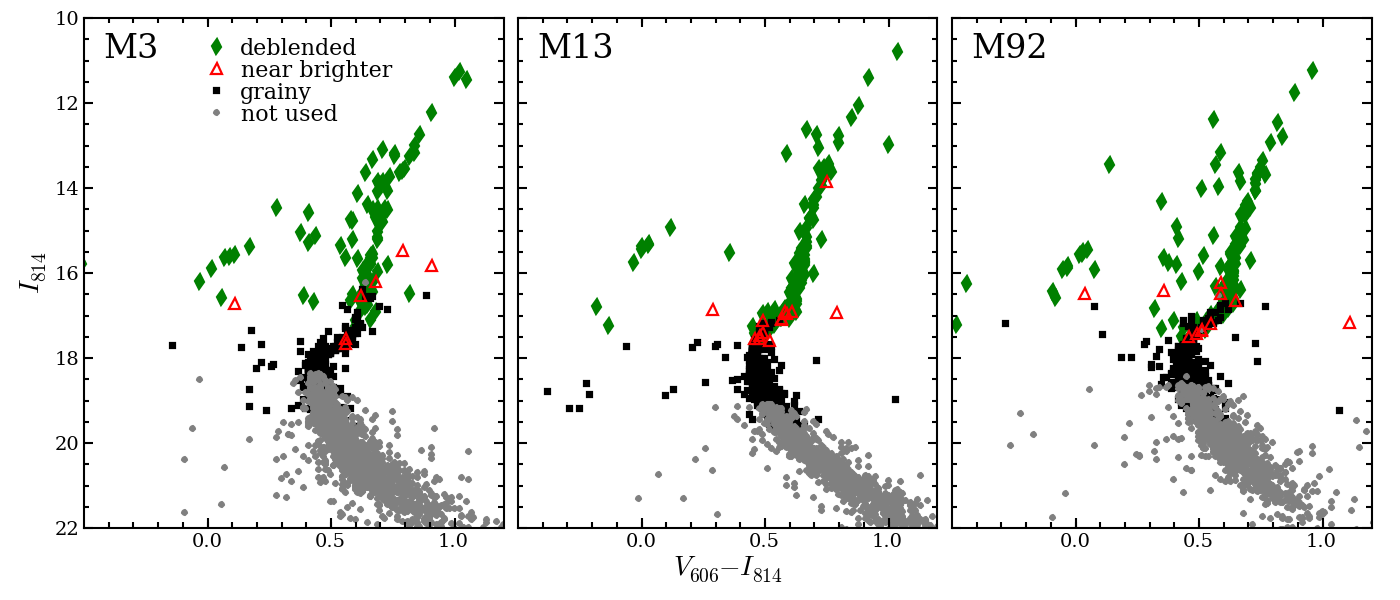}
  \caption{Colour magnitude diagrams of the areas covered by our PMAS observations, coding source categories by different colours and symbols: Individually deblended sources (green diamonds), sources deblended together with a brighter neighbour (red open triangles), sources contributing to the grainy stellar background (black squares), and unused sources (grey circles).}
  \label{fig:kinematic:selection_cmds}
\end{figure*}

\subsection{Deblending spectra in the PMAS data}
\label{sec:kinematic:pmas_analysis}

In \citetalias{2013A&A...549A..71K}, we presented an algorithm to deblend stellar spectra from IFS datacubes via PSF fitting. Starting from an inventory of the stars in the field (hereafter called ``reference catalogue''), we locate the sources in the datacube and determine the subset of sources for which spectra can be deblended at the (typically lower) spatial resolution of the IFS data. Using an iterative approach, we then determine the wavelength dependent PSF as well as the precise coordinate transformation from the reference catalogue to the datacube. In the final step, all the spectra of the resolved sources are simultaneously deblended using a linear least-squares fit. More details are given in \citetalias{2013A&A...549A..71K}.

\subsubsection{Source selection}

First, we determined the expected confusion limit in each datacube, i.e., the magnitude below which no individual sources can be deblended any more. It is roughly equal to the magnitude when the density of brighter sources approaches 0.4 stars per resolution element. Due to the variable seeing conditions, this limit varied from frame to frame. We then determined the expected signal-to-noise ratio (S/N) of each target above the confusion limit and only accepted stars above a S/N threshold of 5. Note that the S/N is an estimate purely based on the brightness of the star, its location in- or outside the PMAS field of view, and the presence of bright neighbours. The actual S/N of each extracted spectrum was determined afterwards as outlined below.

Each spaxel will not only contain contributions from the resolved stars, but also from the unresolved ones and the night sky. We assume that the night sky brightness is constant across the small PMAS field of view and account for it by including an additional component in the analysis of each data cube that is spatially flat. To account for the contribution of the unresolved stars, we considered the following. Stars just below the confusion limit still cause a grainy structure across the field of view, similar to the surface brightness fluctuations observed in nearby galaxies. Therefore, we used an additional component in the deblending process, containing all stars within 2 magnitudes below the confusion limit. The flux ratios of the individual stars in that component were fixed to their relative $I$-band brightnesses. The range of 2 magnitudes was chosen because  stars inside this range account for the vast majority of the graininess and the contribution of even fainter stars is essentially flat across the field of view, so that is is indistinguishable from the contribution of the night sky.

Note that the the inclusion of these unresolved components in combination with the PSF-based extraction of the spectra of the resolved stars accurately cleans the extracted spectra from any (stellar or telluric) background. This is the big advantage of our approach with respect to simpler analysis methods such as aperture extraction, where such a decomposition is not possible.

To visualize the results of the source selection process, we show a colour magnitude diagram of the covered area in each cluster in Fig.~\ref{fig:kinematic:selection_cmds} and highlight the sources that passed the various selection criteria. In all three clusters, our data allowed us to deblend spectra almost all the way down the red giant branch. Note that the few bright sources marked by red triangles in Fig.~\ref{fig:kinematic:selection_cmds} fell below the S/N threshold because of their small distances ($\leq0.3\times$ the seeing FWHM) to a brighter neighbour. For those sources we did not try to deblend individual spectra but accounted for their contributions by modifying the PSF of the neighbour as outlined in \citetalias{2013A&A...549A..71K}. The transitions between the individual selection categories in Fig.~\ref{fig:kinematic:selection_cmds} are not sharp because the confusion limit varied with the observing conditions.

\subsubsection{PSF}

\begin{figure}
 \resizebox{\hsize}{!}{\includegraphics{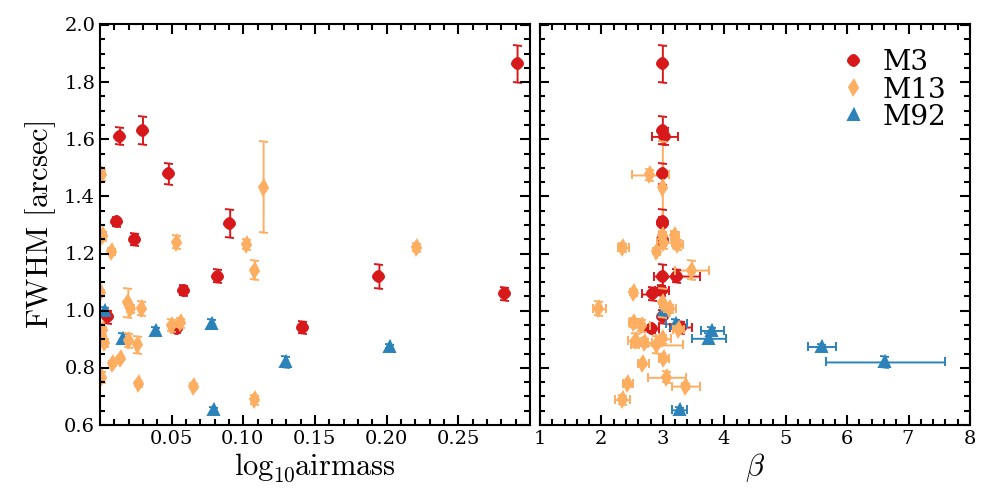}}
 \caption{Comparison of derived PSF parameters. Each panel shows the FWHM: As a function of the airmass of the observation (left), and as a function of $\beta$ (right). Each plotted point corresponds to the mean value obtained in a datacube. Different symbols indicate different clusters as indicated in the legend. The error bars represent the standard deviations over the cubes, including variations with wavelength.}
 \label{fig:kinematic:psf_parameters}
\end{figure}

Owing to the small field of view of PMAS, the data cubes generally did not contain any bright and isolated stars useful for determining the PSF. Therefore, we determined the PSF using all the resolved stars instead. For each such star, a PSF model was centred on the location of the star in the data cube and the model was optimized via a least squares fit to all resolved sources. We modelled the PSF as a Moffat function with up to 4 free parameters: the FWHM, the $\beta$ parameter controlling the kurtosis of the profile, the position angle, and the ellipticity. For each observed datacube, the FWHM was a free parameter in the initial fit, using a prior guess matched to the seeing of the observation. In the majority of the cases, we also allowed $\beta$ to vary. Exceptions were made for some observations where either no reasonably bright star was in the observed field or the seeing was rather poor ($\gtrsim2\arcsec$). In those cases, the contrast was not high enough to yield useful constraints on the value of $\beta$ and we fixed it to a typical value for the observations taken during the same night. Fig.~\ref{fig:kinematic:psf_parameters} shows a summary of all the values we obtained for FWHM and $\beta$. The wide distribution of FWHM values gives a good impression of the different seeing conditions (and thus spatial resolutions) in our data. As a side note, no correlation is observed between the FWHM and the airmass in the observation although one would expect the seeing to increase with airmass. Also, we observe no correlation between the FWHM and $\beta$ and in most cases, the values of the latter scatter around $\beta\sim3$. The rather low values of $\beta$ indicate that the PMAS PSF has extended wings and that a Gaussian profile would not be a valid description of the PSF. There are two notable exceptions with values of $\beta\sim6$. These two datacubes of \object{M92} were observed within the same night in the October 2011 run. During this run we observed only one additional datacube, also of \object{M92}, where we find a rather normal $\beta$ value. So it remains unclear what caused the high values in the two mentioned cases.
Using an elliptical PSF did not significantly improve the results, so we always assumed a circular PSF.

\subsubsection{Source coordinates}

\begin{figure}
 \resizebox{\hsize}{!}{\includegraphics{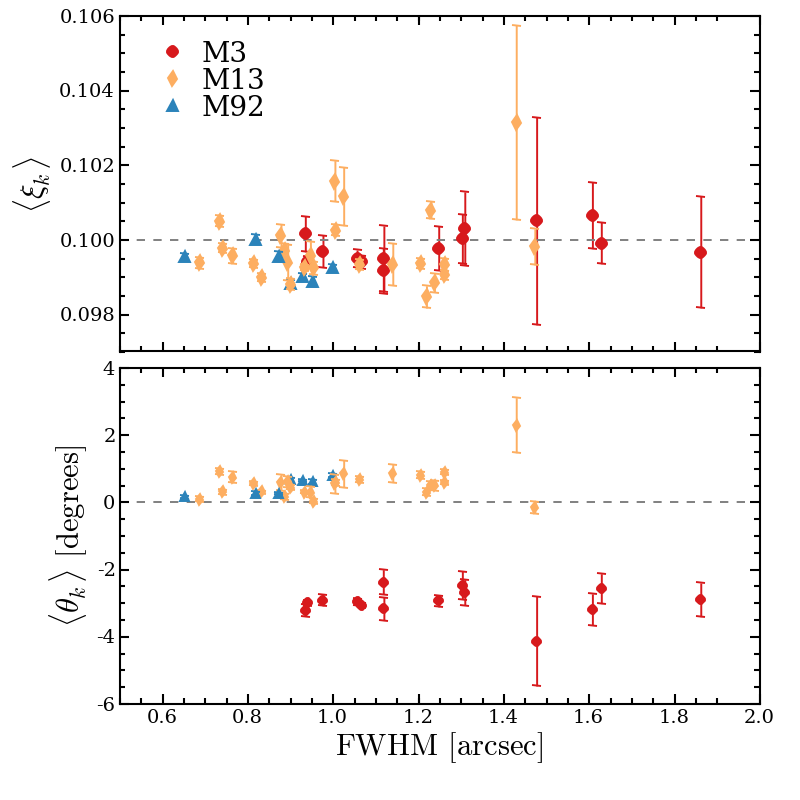}}
 \caption{Average spatial scaling factors (top) and rotation angles (bottom) of all PMAS datacubes with respect to the reference catalogue. As in Fig.~\ref{fig:kinematic:psf_parameters}, the error bars represent the standard deviations over the cubes, including variations with wavelength. Dashed lines in both panels show the values expected for the nominal characteristics of PMAS in the instrument setup that was used.} 
 \label{fig:kinematic:position_parameters}
\end{figure}

We always modelled the coordinate transformation using four free parameters, two to define the scaling and rotation of the reference coordinates into a PMAS field of view, and another two to describe shifts along the $x$- and $y$-axes. PMAS has a nominal spaxel scale of $0\farcs5$ and a fixed rotation angle of zero degrees. The reference coordinates have a pixel scale of $0\farcs05$ with the $y$-axis pointing north \citep{2008AJ....135.2055A}, which is why we expect a scaling factor of $\xi=0.1$ and zero rotation angle. As can be seen from Fig.~\ref{fig:kinematic:position_parameters}, where we show the scaling factors and rotation angles for all datacubes, this is not the case: On average, we obtain $\xi<0.1$, indicating a slightly bigger size of the PMAS spaxels. Simply taking the mean and standard deviation of all measurements gives $\xi=0.0997\pm0.0002$. If we assume that the coordinates are correct in the centre of the PMAS field of view, this corresponds to an uncertainty near the edges of the field of view of $\sim$3\% per spaxel, similar to what we found in our analysis of simulated PMAS data in \citetalias{2013A&A...549A..71K}.

We note that for the simulated PMAS data we also found a mild tendency that the recovered scaling factors came out too large. Compared to the nominal spaxel scale of PMAS, the observed data show the opposite trend. However, differences in the scaling of this order can also occur within the instrument, e.g., caused by a temperature dependent focus. In our dataset, we do not find a significant correlation between the recovered scaling factor and the ambient temperature during the observation. Given the small spread in $\xi$, this is not surprising.

The rotation angle shows significant offsets from zero, up to a few degrees. We observe a bimodal distribution, with the values for the \object{M3} data being consistent with an angle of $\theta=-3\degr$, while for \object{M13} and \object{M92}, the values cluster around $\theta=+0.5\degr$. An explanation for this behaviour is that the \object{M3} data were all obtained in 2010, before the telescope experienced a significant downtime and refurbishment, while the other two targets were observed after that period. As the Cassegrain flange has an adjustable rotation mechanism, the most likely explanation is that during this period of time, when the instrument was dismounted from the telescope, the previous angle setting was lost.

\subsubsection{Extraction of spectra}

\begin{figure*}
 \centering
  \includegraphics[height=17cm,angle=90]{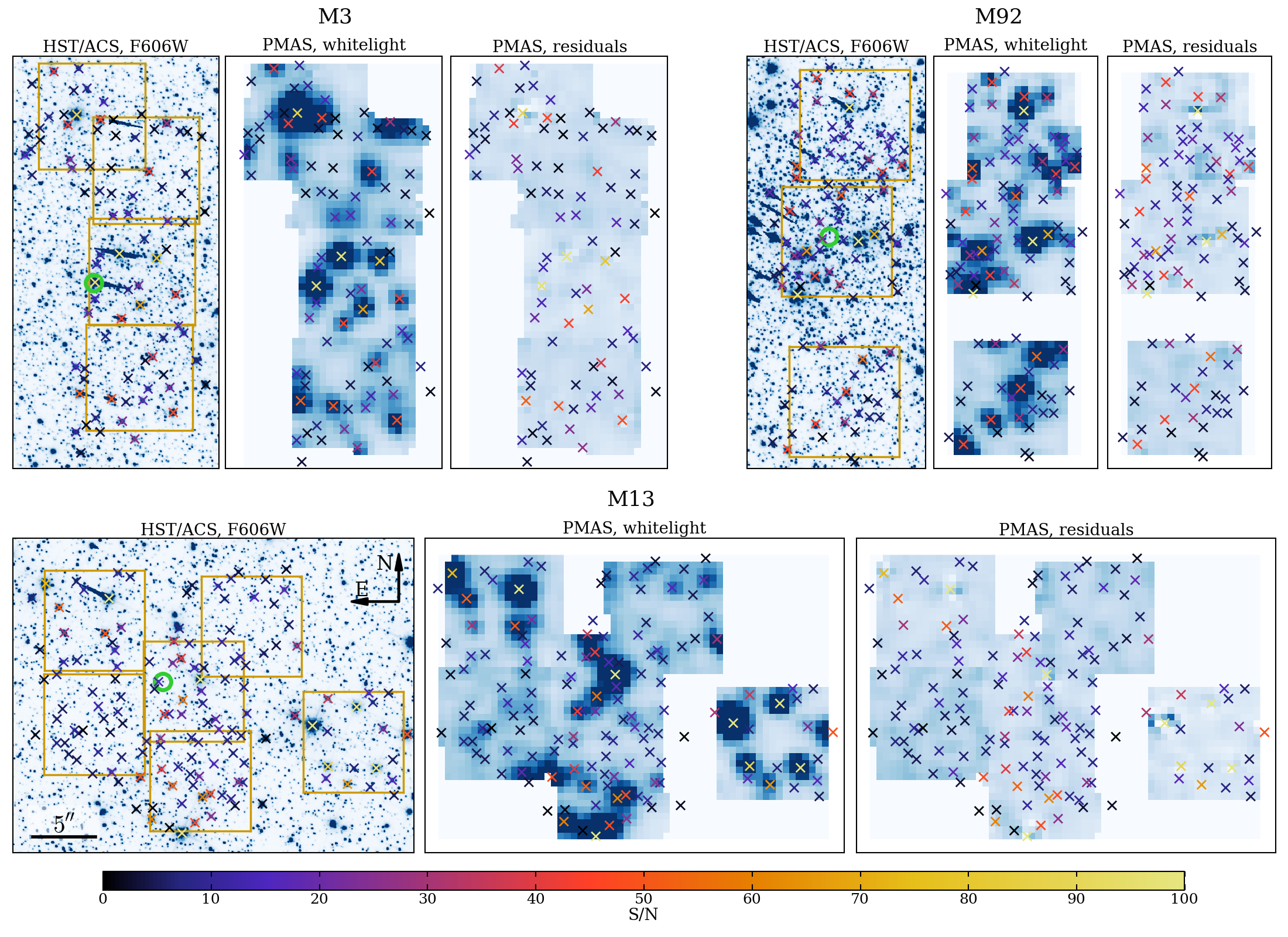}
  \caption{Results of the deblending of the stellar spectra. For each cluster the panels show, from left to right: an HST image of the central region with the PMAS pointings overplotted, a white light image of the combined PMAS data, and the residuals from the PMAS data after the sources were subtracted. In each HST panel, a green circle marks the cluster centre. We highlight the positions of the extracted spectra by coloured crosses, with the colour coding of the crosses matched to the S/N of the respective spectrum as indicated by the colour bar below the panels.}
  \label{fig:kinematic:deblending_results}
\end{figure*}

Using the wavelength dependent PSF parameters and coordinate transformations, we deblended the spectra for the sources identified in the reference catalogue. In Fig.~\ref{fig:kinematic:deblending_results} we present the results of this process for the three clusters. Together with an HST image and a white light mosaic of the integral field data, we also show the residuals after subtraction of the modelled sources. Note that the non-trivial mosaic patterns that we covered with the PMAS observations have two reasons. We already mentioned the pointing inaccuracy of $1-2\arcsec$ that was a consequence of the telescope acquisition procedure. Additionally, some fields were observed because they contained several bright stars and thus allowed for a good PSF reconstruction. One example of such a field is the isolated one to the south west in \object{M13}. The reason that in \object{M3} our pointings are asymmetric with respect to the cluster centre is that the centre we assumed during the observations was offset from the one we used in the analysis. One field in \object{M3} is not shown because it is located at a larger distance ($\sim30\arcsec$) to the cluster centre.

\begin{figure*}
 \centering
 \includegraphics[width=17cm]{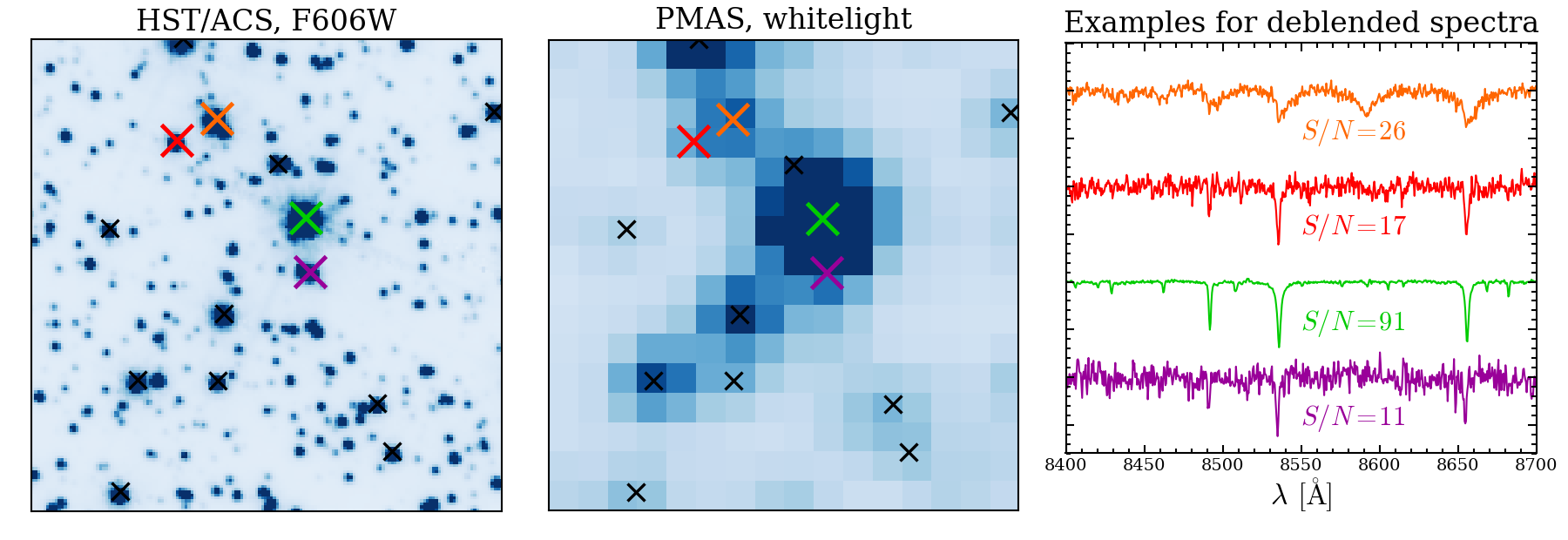}
 \caption{Example spectra deblended from the central PMAS pointing in \object{M13}. The left panel shows an HST image of the region, a white light image of our PMAS data is shown in the central panel. In both panels, black crosses indicate the positions where single stellar spectra were extracted. The spectra of the four sources marked by coloured crosses are depicted in the right panel, the calculated S/N is given below each spectrum. The colour coding used to plot the spectra and to indicate their positions is the same in all three panels.}
 \label{fig:kinematic:m13_example_spectra}
\end{figure*}

In each panel in Fig~\ref{fig:kinematic:deblending_results}, coloured crosses indicate the positions where stellar spectra were extracted, with the colour coding matched to the S/N of the spectrum. The S/N values were determined following the prescription of \citet{2007STECF..42....4S}. The number of stellar spectra that we deblended in such a small region of the sky is quite remarkable. As expected, the spectra cover a broad range in S/N, with the brightest stars having values of $\mathrm{S/N}>100$. Some example spectra, extracted from the central pointing in \object{M13}, are shown in Fig.~\ref{fig:kinematic:m13_example_spectra} together with their positions in the cluster. For the brightest giant near the centre of the field of view we obtain a spectrum with a very high S/N as expected. The spectrum of the star uncovered from under the PSF wings of the bright star has a substantially lower S/N. Note that PSF fitting is the only technique that allows one to obtain an uncontaminated spectrum for this star at all. The same is true for stars of similar brightnesses close to each other. This is illustrated by the two stars towards the upper left corner of the field of view. As can be seen from the right panel of Fig.~\ref{fig:kinematic:m13_example_spectra}, one of them is a horizontal branch star that has a spectrum that is different from the one of its neighbour, with no mutual contamination visible in the spectrum of either star. This highlights the power of our approach. In a few cases (marked by open triangles in Fig.~\ref{fig:kinematic:selection_cmds}) the distance between two stars is so small that even with PSF fitting techniques they cannot be accurately deblended because their PSF images become indistinguishable. In such cases we obtained a single spectrum for the two stars that was accordingly flagged during the further analysis.
In total, we extracted spectra for 102 stars in \object{M3}, 141 in \object{M13}, and 98 in \object{M92}. The subsequent analysis of these extracted spectra will be discussed below.

When comparing the results for \object{M92} with those obtained for \object{M13} in Fig.~\ref{fig:kinematic:deblending_results} it becomes clear how strongly the seeing influences the quality of the data. Relative to the number of pointings, we can extract significantly more stars in the \object{M92} data than in the \object{M13} data. On average, the stars in \object{M92} also have a higher S/N than those in \object{M13}. The observations of \object{M3} were carried out under similar conditions to those of \object{M13}, but the average S/N of the stars is higher. The reason for this can be seen in Fig.~\ref{fig:sample:mag_vs_density}, where we plot the stellar density as a function of magnitude. The density of \object{M13} increases much steeper towards the confusion limit compared to \object{M3} or \object{M92}. This implies that in \object{M13} we select many stars of almost equal brightness right above the confusion limit.

\subsection{Radial velocities}
\label{sec:kinematic:rv}

\subsubsection{Cross-correlation of PMAS spectra}

The determination of radial velocities via cross-correlation requires an accurate determination and subtraction of the continuum of each spectrum. To do so, we followed \citet{2011AJ....141..187S}: Each spectrum was initially fitted with a polynomial, then $\kappa$-$\sigma$-clipping was performed on the residuals, using asymmetric $\kappa$-values for the upper and the lower threshold. While the upper threshold remained fixed at $\kappa_\text{up}=3$, the lower threshold $\kappa_\text{low}$ was adjusted based on the S/N of the spectrum, with higher values being used for lower S/N values and vice versa. This was necessary because spectral lines in high-S/N spectra must be clipped, whereas noise spikes in low-S/N spectra should not be spuriously identified as spectral lines.

We cross-correlated the extracted spectra against synthetic templates, taken from the library of \citet{2005A&A...442.1127M}, to determine absolute velocities. A potential source of error when using synthetic spectra is that their line profiles do not match the line spread function (LSF) of the observations. The PMAS LSF, as measured from the emission lines of the arc lamp calibration data, showed strong wavelength-depended asymmetries. To account for those, we fitted each emission line with a Gauss-Hermite function and modelled the change of the shape parameters with wavelength as a low order polynomial. This resulted in a wavelength-dependent model of the LSF that each template was convolved with before the cross-correlation. The correction for the shape of the LSF was small ($\lesssim1\,\mathrm{km\,s^{-1}}$) in \object{M13} and \object{M92}, yet significant ($\sim4\,\mathrm{km\,s^{-1}}$) for the \object{M3} data, observed in a single run in 2010.

Our samples not only include red giants, but also stars on the horizontal branch. As visible from Fig.~\ref{fig:kinematic:m13_example_spectra}, these stars have broad Paschen absorption lines in the covered spectral range. To obtain a matching template for each extracted star, we constructed an isochrone for each cluster using the tool of \citet{2008A&A...482..883M} and inferred effective temperatures $T_\mathrm{eff}$ and surface gravities $\log g$ for the cluster stars. Each spectrum was then cross-correlated against the library template that best matched the star in terms of $T_\mathrm{eff}$, $\log g$, and $[\mathrm{Fe/H}]$.

To assess the quality of a cross-correlation result, we used the $r_\mathrm{cc}$ statistic for a cross correlation peak \citep{1979AJ.....84.1511T},
\begin{equation}
 r_\mathrm{cc} = \frac{h}{\sqrt{2}\sigma_\mathrm{a}}\,,
\label{eq:rv:rcc_statistics}
\end{equation}
where $\sigma_\mathrm{a}$ is the normalized standard deviation of the antisymmetric component of the cross correlation spectrum. \citeauthor{1979AJ.....84.1511T} showed that the uncertainty tailored to a measured radial velocity is proportional to $(1+r_\mathrm{cc})^{-1}$. A common method to determine the proportionality factor $C$ is to via the repeated observations of the same stars \citep[e.g.][]{1988AJ.....96..123P,2007AJ....133.1041D}, as the quantity
\begin{equation}
 \frac{v_{\mathrm{rad},2}-v_{\mathrm{rad},1}}{\left((1+r_{\mathrm{cc},1})^{-2}+(1+r_{\mathrm{cc},2})^{-2}\right)^{1/2}}\,,
\end{equation}
should be distributed normally with a standard deviation of $C$. However, as \citet{1998PASP..110..934K} showed the uncertainty should also be proportional to the width $w$ of the correlation peak. As a consequence of the different line widths in our observed spectra (cf. Fig.~\ref{fig:kinematic:m13_example_spectra}), we obtained a large range in $w$ values. For this reason, we investigated the scatter of the quantity
\begin{equation}
 \frac{v_{\mathrm{rad},2}-v_{\mathrm{rad},1}}{\left(\left(\frac{w_1}{1+r_{\mathrm{cc},1}}\right)^{2}+\left(\frac{w_2}{1+{r_{\mathrm{cc},2}}}\right)^{2}\right)^{1/2}}\,,
\end{equation}
and found it to be normally distributed with standard deviation of $\tilde{C}=0.31$, so our final equation for obtaining uncertainties was
\begin{equation}
 \epsilon_{\mathrm{v_{rad}}} = \frac{0.31\cdot w}{1+r_\mathrm{cc}}\,.
\label{eq:rv:uncertainty}
\end{equation}

\begin{figure*}
 \centering
 \includegraphics[width=18cm]{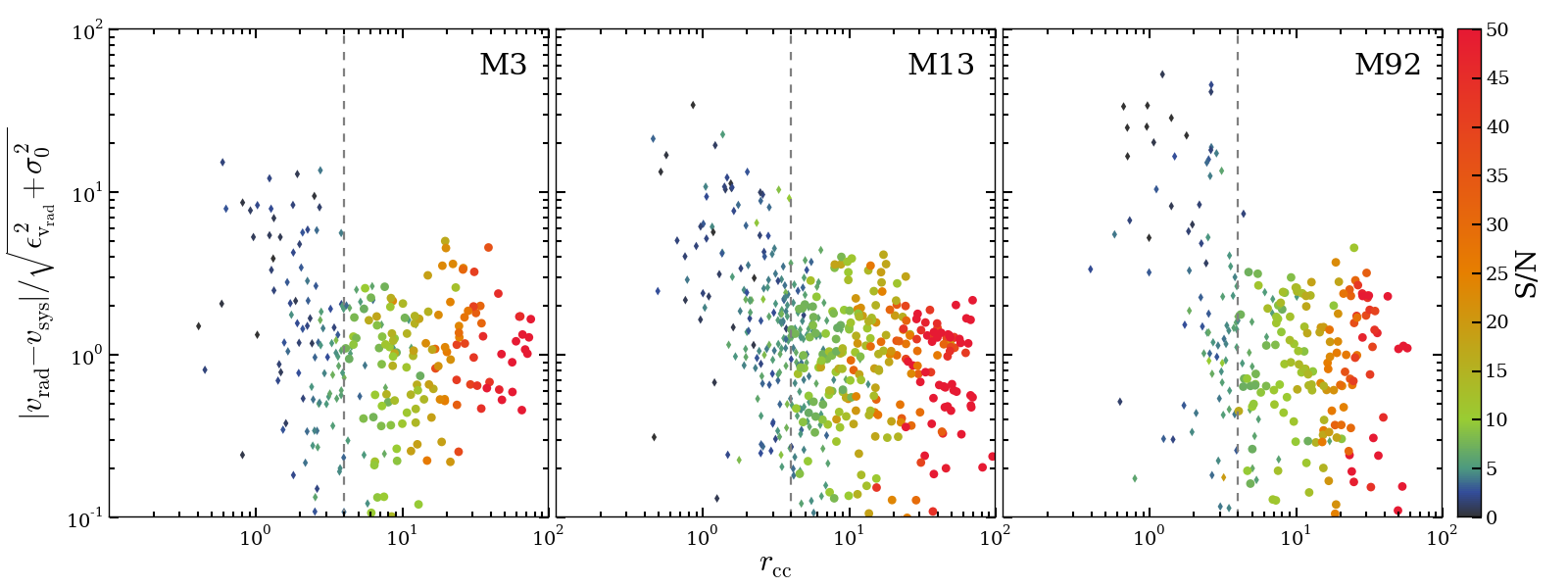}
 \caption{Offsets between the measured radial velocities and the systemic velocity of each cluster as a function of the cross correlation statistic $r_\mathrm{cc}$. The offsets are normalized using the measurement uncertainties and the central velocity dispersion of each cluster (cf. Table \ref{tab:pmas_clusters}). Data are colour coded according to the S/N of the individual spectra as indicated by the colour bar to the right-hand side of the plot. To highlight differences at low S/N values, the colour coding is truncated at $\mathrm{S/N}=50$. Spectra below the cut value of $\mathrm{S/N}=7$ are plotted with small squares, those above with larger circles. The vertical lines denote the selected cut in $r_\mathrm{cc}$.}
 \label{fig:kinematic:blunder_diagram}
\end{figure*}

In Fig.~\ref{fig:kinematic:blunder_diagram} we plot the distribution of our measured radial velocities for the three clusters in the $r_\mathrm{cc}$-$v_\mathrm{rad}$ plane. \citet{1998PASP..110..934K} called it the ``blunder diagram'' and proposed to use it in order to set a reliability threshold on $r_\mathrm{cc}$ in the sense that for lower values of $r_\mathrm{cc}$, the measured radial velocities can be arbitrarily high. \citet{2002AJ....124.3270G} showed that a threshold in $r_\mathrm{cc}$ in combination with a cut in the S/N of the sources is an efficient way to clean a radial velocity sample from unreliable measurements. In our case, a combination of $r_\mathrm{cc, min}=4$ and $\mathrm{S/N}_\mathrm{min}=7$ proved to be a good choice. After applying these selection criteria, we were left with 137 reliable velocity measurements of 56 stars in \object{M3}, 262 measurements of 86 stars in \object{M13}, and 165 measurements of 83 stars in \object{M92}. They are summarized in Tables~D1 (\object{M3}), D2 (\object{M13}), and D3 (\object{M92}), available in the online version of this article.

\subsubsection{Literature data}

\begin{table*}
 \caption{Literature data used to complement our radial velocities}
 \label{tab:kinematic:literature_studies}
 \centering
 \begin{tabular}{ c c l c l }
  \hline\hline
   Reference & Abbr. & Telescope(s)/Instrument(s) & Resolution\tablefootmark{a} & Targets ($n_\mathrm{star}$) \\ \hline
  \citet{1979AJ.....84..752G} & GG79 & Palomar/radial velocity spectrometer & $1\,\mathrm{km/s}$ & \object{M3} (111) \\
  \citet{1987AJ.....93.1114L} & L+87 & Palomar/radial velocity spectrometer & $1\,\mathrm{km/s}$ & \object{M13} (147) \\
  \citet{1988AJ.....96..123P} & P+88 & MMT, Tillinghast 1.5m/echelle spectr. & $1\,\mathrm{km/s}$ & \object{M3} (111) \\
  \citet{1999PASP..111.1233S} & S+99 & Mayall 4m, WIYN/Hydra & $\geq$11\,000 & \object{M3} (87), \object{M13} (150), \object{M92} (35)\tablefootmark{b} \\
  \citet{2000AJ....119.2895P} & P+00 & WIYN/Hydra & 14\,000 & \object{M3} (77), \object{M13} (78), \object{M92} (61) \\
  \citet{2007AJ....133.1041D} & D+07 & WIYN/Hydra & 14\,000 & \object{M92} (306) \\
  \citet{2009AJ....137.4282M} & M+09 & MMT/Hectochelle & 34\,000 & \object{M13} (123), \object{M92} (64) \\
  \hline
 \end{tabular}
 \tablefoot{
  For each reference, the observed clusters are given together with number of stars observed per cluster.\newline  
  \tablefoottext{a}{Where available, we provide the spectral resolution $R$, otherwise, the typical uncertainty of a radial velocity measurement is stated.}\newline{}
  \tablefoottext{b}{Excluded, see discussion in App.~\ref{app:literature}.}}
\end{table*}

Our data cover the very central region of each cluster. Clearly, this is where our crowded field IFS approach is most efficient. To probe the kinematics over the full radial extent of our targets we searched the literature for additional radial velocity data. At larger distances to the centres the source densities become significantly lower so that these regions become more accessible also for traditional spectroscopy. When collecting these complementary data from the literature we concentrated on studies that involved more than a minimum of about $50$ stars. We only selected data where meaningful uncertainties were provided with the measured radial velocities. A summary of the catalogues that we included is provided in Table~\ref{tab:kinematic:literature_studies}. All literature studies that we used provide radial velocities with typical uncertainties $\leq1\,\mathrm{km}\mathrm{s}^{-1}$.

To obtain the final set of radial velocities, we combined the results from the individual studies of each cluster. In this process that is described in detail in Appendix \ref{app:literature}, we checked the consistency of the velocities reported by the individual studies and flagged stars that showed evidence for radial velocity variations. We summarize the results of our literature search in Tables~D4 (\object{M3}), D5 (\object{M13}), and D6 (\object{M92}), also available in the online version of this paper.

\subsubsection{Comparison between PMAS and literature data}

\begin{figure}
 \centering
 \resizebox{\hsize}{!}{\includegraphics{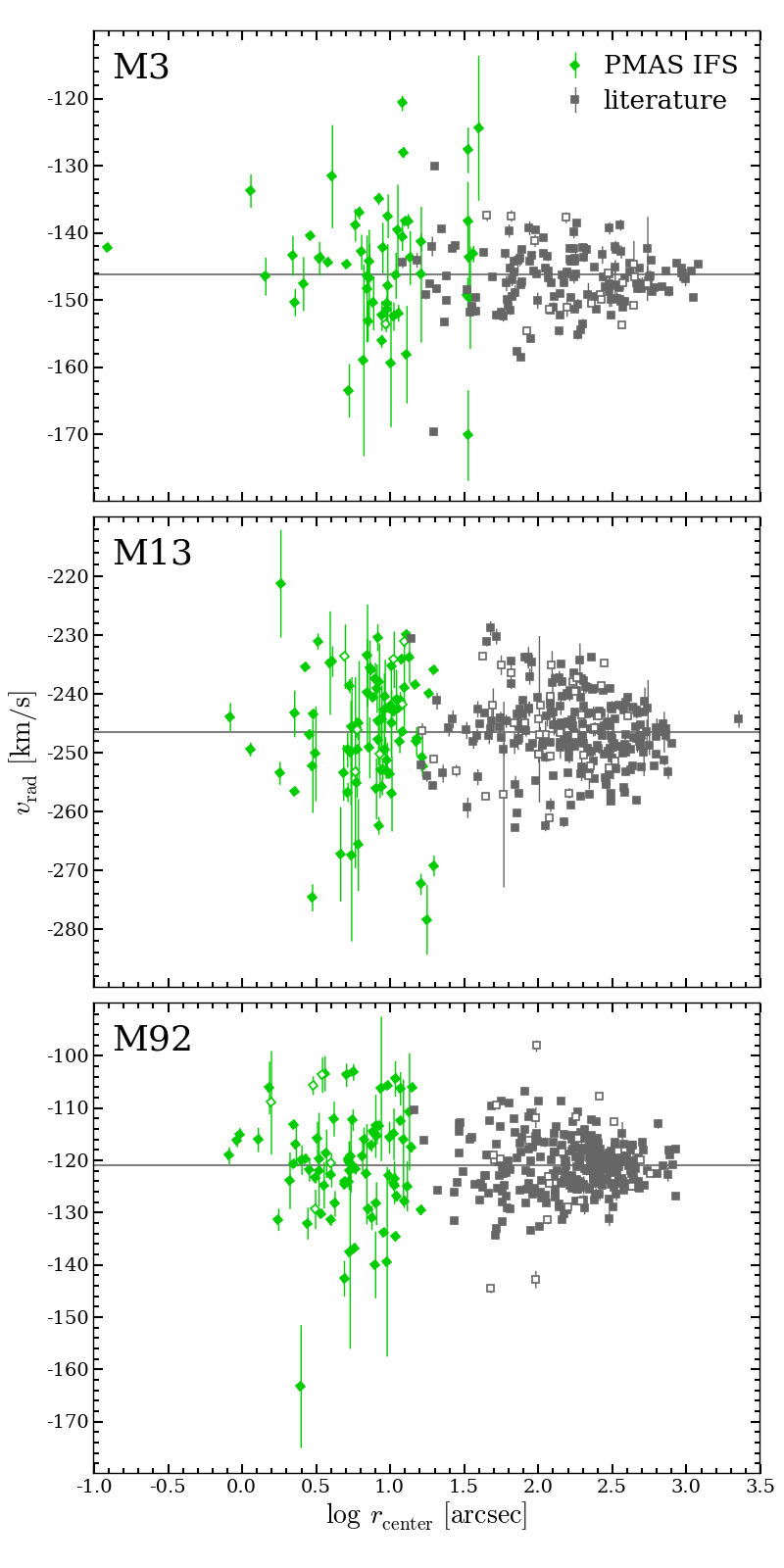}}
 \caption{Radial velocities of all stars in the three clusters as a function of distance to the cluster centre. Radial velocity measurements from our PMAS data are plotted using green diamonds, black squares refer to literature data. Open symbols are used to indicate stars that show variability, either photometrically or kinematically. Horizontal lines indicate mean velocities of the samples. Note that for clarity, $2$ stars in M3 and in M13 with large uncertainties ($>20\mathrm{km\,s^{-1}}$) are not shown.}
 \label{fig:kinematic:distance_vs_velocity}
\end{figure}

In Fig.~\ref{fig:kinematic:distance_vs_velocity} we plot the radial velocities we obtained from our PMAS data and from the literature as a function of the distance to the centre of each of the three clusters. The complementarity is obvious: Traditional spectroscopy is nearly ``blind'' in the central regions of the clusters, where instruments such as PMAS provide the highest gain, limited only by their small field of view.

When comparing the mean velocities of the subsamples, we obtain good agreement for \object{M13} (PMAS: $-246.7\pm1.2\,\mathrm{km\,s^{-1}}$, literature: $-246.4\pm0.4\,\mathrm{km\,s^{-1}}$) and \object{M92} (PMAS: $-120.7\pm1.0\,\mathrm{km\,s^{-1}}$, literature: $-121.1\pm0.3\,\mathrm{km\,s^{-1}}$). Only in the case of \object{M3}, a larger discrepancy of $2.8\,\mathrm{km\,s^{-1}}$ between the mean velocities is observed (PMAS: $-244.2\pm1.2\,\mathrm{km\,s^{-1}}$, literature: $-247.0\pm0.4\,\mathrm{km\,s^{-1}}$). Note that this discrepancy increases if the LSF of PMAS is not accounted for.

For further investigation we matched the literature data to our reference catalogue to check whether some stars were covered by our PMAS footprint. This was done by overlaying both the sources and our reference catalogue objects on an HST footprint and visually identifying the counterparts. In the process of doing this we found that some stars have very inaccurate positional data in the literature, especially in \object{M3}, where we observed an average offset of $\sim1\arcsec$. The scatter in \object{M13} is only $\sim0.2\arcsec$, but some outliers with offsets around $2\arcsec$ were observed. The positions in \object{M92} are the most accurate ones, showing a scatter of $\sim0.1\arcsec$ and only some outliers with offsets $\sim1\arcsec$. A likely explanation for the inaccuracies is that the input catalogues used in the literature studies reach back to, e.g., \citet{1905POPot..50....1L}. Such offsets render the measured radial velocities less reliable. In a crowded stellar field such as a globular cluster, an observation will yield a spectrum even if the target star is (partly) missed. However, this spectrum may not only contain light coming from the target star but also a contribution from the numerous fainter stars in its vicinity. Nearby ($\lesssim1\arcsec$) bright stars can cause a similar effect and are therefore equally undesirable. We flagged all stars that (i) had bright neighbours that likely influenced the measurement of the radial velocity or (ii) were significantly offset from a bright HST counterpart or even had no obvious counterpart on the HST footprint at all. These stars were excluded from further analysis. The flags are also included in Tables~D4, D5, and D6.

In \object{M3} and \object{M13}, a few of the stars with radial velocity measurements available in the literature are located within our PMAS pointings: four in \object{M13} and five in \object{M3}. When comparing our radial velocity measurements with those in the literature, we find an excellent agreement for the sources in \object{M13}, the rms deviation of $0.7\,\mathrm{km\,s^{-1}}$ is the same as the typical measurement uncertainties. However, in \object{M3} the situation is more complex: the rms deviation of $2.6\,\mathrm{km\,s^{-1}}$ exceeds the typical measurement uncertainties by a factor of $4$. Intrinsic RV variability is unlikely to be the reason for this, as none of the stars shows signs of variability, neither in the literature data nor in our dataset. Also, accounting for the measured offset in the mean velocities of the samples does not improve the agreement. We rather suspect that the literature measurements suffer from inaccuracies in the assumed positions of the stars. All stars where we observed a significant discrepancy between our data and the literature were also flagged when checking their positions in the HST image. They were therefore not included in the further analysis.

Inspection of the radial velocities shown in Fig.~\ref{fig:kinematic:distance_vs_velocity} reveals that some stars have high relative velocities compared to the cluster means. The presence of such high-velocity stars in \object{M3} was already noted by \citet{1979AJ.....84..752G}, who detected two stars (those with the highest relative velocities in the literature sample of \object{M3}) that showed velocities that could not be explained by the (Gaussian) velocity distribution of the cluster members. \citet{1979AJ.....84..752G} speculated on their origin and considered it unlikely that they were formed either via binary disruption or strong two-star encounters, suggesting that they might be remainders from the formation of the cluster (although the slowing down time due to dynamical friction is quite short). More recently, \citet{2012A&A...543A..82L} suggested that such a star can be formed in a single encounter between a main sequence binary star and a stellar-mass black hole.

When identifying high-velocity stars, one has to keep in mind that measurement uncertainties will on average push the velocity away from the cluster mean. So the important quantity is not the absolute velocity offset but the offset relative to the corresponding measurement uncertainty. For this reason, we determined for each star an effective velocity dispersion as $\sqrt{\epsilon_\mathrm{v_{rad}}^2+\sigma_\mathrm{0}^2}$, where $\epsilon_\mathrm{v_{rad}}$ is the measurement uncertainty of the star and the central velocity dispersion of the cluster is taken from Table~\ref{tab:pmas_clusters}. We identified as a potential outlier each star with a relative velocity larger than $3\times$ the effective dispersion. In the literature data, except for the stars in \object{M3} already identified by \citet{1979AJ.....84..752G}, only the sample of \object{M92} contains some candidates, yet all three of them show evidence for RV variability. However, our PMAS data contain four suspicious stars for which no RV variability is observed, one in \object{M3}, another two in \object{M13}, and a horizontal branch star with poorly determined radial velocity in \object{M92}. While it seems reasonable to attribute their presence to the mechanisms that we mentioned earlier, we did not want to exclude the possibility that these stars simply represent the tail of the velocity distribution. We will discuss their influence when we calculate dynamical models for the clusters.

We also investigated the possibility that the high velocity stars are not members of the clusters. However, $V$- and $I$-band colours place them nicely on the red giants branches of the respective clusters (and on the horizontal branch of \object{M92}). Further, we determined the strengths of the calcium triplet lines. For red giant stars, the equivalent widths of these lines can be approximated as a linear function of $V$-magnitude and are commonly used as a measure for metallicity \citep[e.g.][]{1991AJ....101.1329A}. For two out of the three candidates for which we have spectra available, the measured equivalent widths agree well with those of other red giants with comparable magnitudes. One candidate in \object{M13} (the star with ID $56100$ in Table~D2) shows an offset of $\sim20\%$ to lower equivalent widths.

\subsection{Variable stars}

As mentioned above, stars with variable radial velocities are flagged in order to exclude them from the analysis of the cluster kinematics. Since radial velocity variations either originate from intrinsic variability or the presence of binary stars, the measured velocities do not necessarily trace the cluster gravitational potential.
Intrinsic variability can easily cause variations of the order of the velocity dispersion of the cluster. For example, an RR Lyrae star with a period of $0.5$ days will show variations by $\pm20\,\mathrm{km\,s^{-1}}$, assuming a change in radius by $0.5$ solar radii \citep{2010A&A...519A..64K}.

The importance of binary stars is expected to decrease with the density and the mass of the cluster. \citet{2008A&A...480..103K} predicted the impact of binaries on the estimate of the dynamical mass of a cluster depending on its mass and half-mass radius. They found that the influence of binaries can be neglected if the central velocity dispersion is $\gtrsim10\,\mathrm{km\,s^{-1}}$ while they dominate the dynamics in the regime $\lesssim1\,\mathrm{km\,s^{-1}}$. This puts our clusters right on the edge of where binaries become negligible.
Observational determinations of the binary fraction in the three clusters were carried out, e.g., by \citet{2012A&A...540A..16M} using HST photometry. They determined low binary fractions of about $1-2\%$. The determination of binary fractions based on spectroscopy is challenging and so far limited to less dense clusters \citep[e.g., \object{M4},][]{2009A&A...493..947S}.

\begin{figure}
 \centering
 \resizebox{\hsize}{!}{\includegraphics{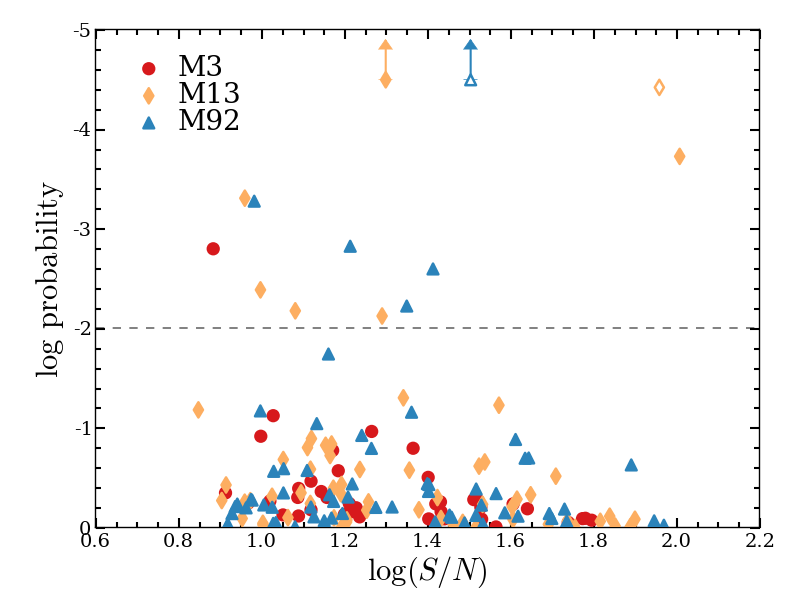}}
 \caption{Time variability in the measured radial velocities in our PMAS data for all stars where multiple radial velocity measurements are available. We show, as a function of the S/N of a star, the probability that the maximum velocity difference is consistent with the measurement uncertainties. A horizontal line indicates the thresholds used to identify variable stars. Stars represented by open symbols are known photometric variables.}
 \label{fig:kinematic:variability_pmas}
\end{figure}

We searched for stars with variable radial velocities in our PMAS data by comparing the radial velocities obtained for the same stars from different exposures. Observations during two epochs only exist for part of the covered fields (5/6 observed fields in \object{M13}; 1/3 fields in \object{M92}), yet pulsating stars or close binaries can show strong velocity variations during a single night. Therefore we plot in Fig.~\ref{fig:kinematic:variability_pmas} the maximum measured differences in the radial velocities for all observed stars where at least two reliable measurements are available. We used a probability $<1\%$ that the offset is consistent with the measurement uncertainties as a criterion for a variable star. This results in 1 RV variable out of 40 stars in \object{M3}, 7 out of 61 in \object{M13}, and 5 out of 62 in \object{M92}.

It is quite remarkable that the two stars with the highest measured S/N in M13 (with ID 58408 and 55692, respectively, in our reference catalogue) are both flagged as possible RV variable stars. They are also the brightest stars in the entire stellar sample of our PMAS data of \object{M13}, both lying at the tip of the red giant branch. \citet{1979AJ.....84..752G} already noticed that such stars often show stronger variability than expected based on the measurement uncertainties. They concluded that likely the stars experience a low magnitude pulsation and added an additional uncertainty of $0.8\,\mathrm{km\,s^{-1}}$ to account for this ``jitter''. The fainter of the two stars is also included in the compilation of variable stars in globular clusters by \citet{2001AJ....122.2587C} as a pulsating star.

Two other stars identified in the catalogue of \citet{2001AJ....122.2587C} as RR Lyrae stars were covered by our PMAS data, one in \object{M92} and one in \object{M3}. For the star in \object{M92} we observe a change in the radial velocity of $\sim10\,\mathrm{km\,s^{-1}}$ during a single night, while the object in \object{M3} does not show significant variations in our observations. However, we flagged the two stars accordingly in Tables~D1 and D3 and excluded them from further analysis.

In the literature data we find evidence for variable radial velocities in 16 out of 142 stars with multi-epoch data in \object{M3}, 4 of which are known photometric variables.  In \object{M13}, the numbers are 31 out of 200 with 11 photometric variables and in \object{M92}, 20 out of 250 with 2 photometric variables included. Combined with our PMAS data, the fraction of RV variable stars we detect is $9.3\%$ in \object{M3}, $14.5\%$ in \object{M13}, and $8.0\%$ in \object{M92}. These are higher than the photometric binary fractions determined by \citet{2012A&A...540A..16M} in the core of each cluster ($\sim1\%$). While these fractions were obtained using main sequence stars, we targeted giant stars were pulsating stars are more common. We can conclude that, although some RV variable stars in our samples may remain undetected because of our sparse sampling of the temporal domain, both, our integral field data and the literature data are consistent with low binary fractions.

\section{Cluster dynamics}
\label{sec:dynamics}

The radial velocities measured from the PMAS data have uncertainties ranging from $\lesssim1\ \mathrm{km\,s^{-1}}$, well below the expected velocity dispersions of the clusters, to $\gtrsim10\ \mathrm{km\,s^{-1}}$, comparable to the expected dispersions. Large uncertainties, however, do not imply that the measurements are useless in the determination of the velocity dispersion. An efficient way to handle the uncertainties in the analysis of the cluster dynamics was summarized by \citet{1993ASPC...50..357P} who used a maximum likelihood approach for this task. The relevant formulae to obtain $v_\mathrm{sys}$ and $\sigma_\mathrm{los}$ --- the mean velocity and the velocity dispersion along the line of sight --- from a sample of velocity measurements are provided in \citet{1993ASPC...50..357P} and not repeated here.

To obtain uncertainties for the maximum likelihood estimates, we used Monte Carlo simulations: 1000 realizations were created for each stellar sample by assigning each star in the sample a new velocity drawn from from a velocity distribution with effective dispersion $\sqrt{\epsilon_\mathrm{v_{rad}}^2+\sigma_\mathrm{los}^2}$, where again $\epsilon_\mathrm{v_{rad}}$ is the velocity measurement uncertainty of the star, and re-calculating $\sigma_\mathrm{los}$. A convenient side effect of these simulations was that they also allowed us to check for any bias in the maximum likelihood estimators which are known to become increasingly biased when the number of stars decreases or the measurement uncertainties grow relative to $\sigma_\mathrm{los}$. We found that in general the velocity dispersions are slightly underestimated, but the offsets were always below $1~\mathrm{km\,s^{-1}}$ and well within the respective error bars. Not surprisingly, the bias was stronger for our PMAS data ($\sim0.5\ \mathrm{km\,s^{-1}}$) than for the literature data ($\sim0.2\ \mathrm{km\,s^{-1}}$). This is a consequence of the wide range of velocity uncertainties in our data.

Another way to obtain the uncertainties would be via a jackknife technique \citep{1993stp..book.....L}, where the velocity dispersion is re-calculated leaving out one star at a time. The agreement with our simulations is in most cases better than $0.2\mathrm{km\,s^{-1}}$. However, the jackknife technique is only valid if the data points have comparable uncertainty variances \citep[e.g.][]{2010arXiv1008.4686H} which is not strictly the case for our data set. Therefore we stick to the results from the Monte Carlo simulation.

\subsection{Rotation in the clusters}
\label{sec:dynamics:rotation}

\begin{figure*}
 \centering \includegraphics[width=17cm]{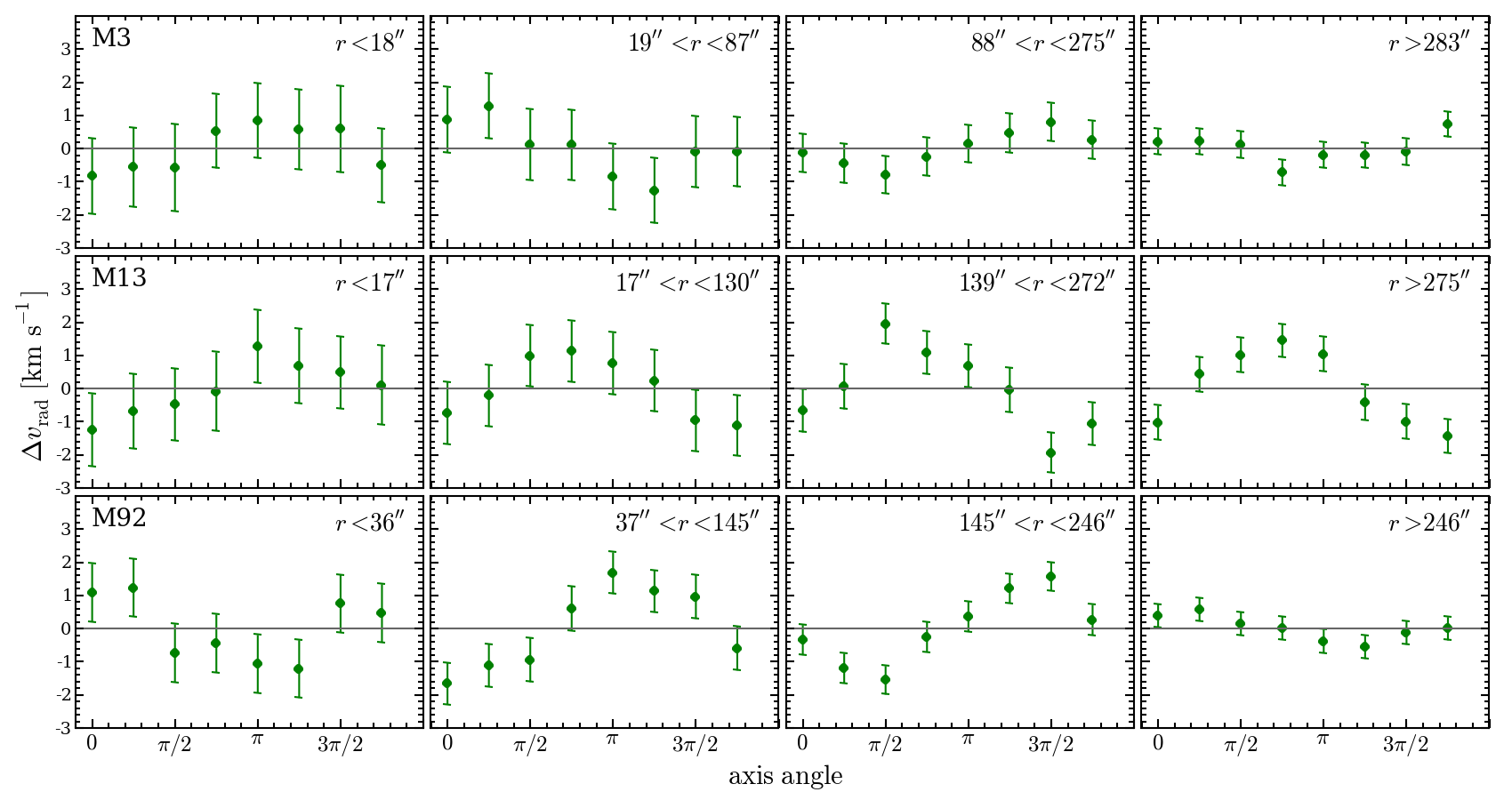}
 \caption{Search for internal rotation of the clusters \object{M3} (top), \object{M13} (center), and \object{M92} (bottom). For each cluster, 4 radial annuli around the cluster centre are shown, with radius increasing from left to right. The stars in each bin were divided into two subsamples under the assumption of different rotation axes and the difference in mean velocity was calculated. The position angle is zero when aligned in east-west direction and increases anticlockwise. The extent of the radial bins is indicated in each panel.}
 \label{fig:dynamics:binned_rotation}
\end{figure*}

We first checked if any of the clusters in our sample shows signs of rotation. To this aim we separated our data into 4 radial bins per cluster. In each bin the measured radial velocities were further divided into two halves with respect to an assumed rotation axis. We determined the mean velocity and its uncertainty in each half-bin using the maximum likelihood approach and determined the difference in mean velocity to the whole sample. This step was repeated assuming different rotation angles. In case of a non-zero rotational component, one expects a sinusoidal variation of the mean velocity difference with axis angle. The relation between the two quantities in the three clusters is shown in Fig.~\ref{fig:dynamics:binned_rotation}.

In \object{M3}, only a weak rotation signal with projected amplitude $\lesssim1\mathrm{km\,s^{-1}}$ is observed. For \object{M13}, \citet{1987AJ.....93.1114L} already reported that this cluster shows rotation. While no clear rotational signal is detected in the central bin, the three outer bins suggest rotation with a projected amplitude $\sim2\ \mathrm{km\,s^{-1}}$, in reasonable agreement with the measurements of \citeauthor{1987AJ.....93.1114L}. The rotation axis is tilted $\sim45^{\circ}$ with respect to the east-west direction. Finally, in \object{M92} all four radial bins suggest a small rotational component, with a projected amplitude $\sim1\text{--}2\ \mathrm{km\,s^{-1}}$. Yet, when comparing the curves depicted in the different panels shown for \object{M92} in Fig.~\ref{fig:dynamics:binned_rotation}, the axis angle suggested by the data seems to be quite variable.

In summary, all clusters display signs of internal rotation. However, unless very unfortunate projection effects are at work, its amplitudes are much lower than the expected velocity dispersion. This is important because the Jeans models we discuss below assume spherical symmetry which would be broken in strongly rotating stellar systems. The generalisation to such systems can be done via axis-symmetric models that we do not consider here.

A word of caution is in order regarding \object{M3}. As discussed in Sect.~\ref{sec:kinematic:pmas_analysis}, our PMAS pointings only cover the region west of the cluster centre, so in the unfortunate case that the cluster rotates around the north-south axis, we would not detect the rotation signal. Such rotation might serve as an explanation for the larger discrepancy in the mean velocities of our PMAS data and the literature sample (cf. Sect.~\ref{sec:kinematic:rv}). However, the offset is still significantly smaller than the velocity dispersion expected in the centre.

\subsection{Velocity dispersion profiles}

\begin{figure*}
 \centering\includegraphics[width=17cm]{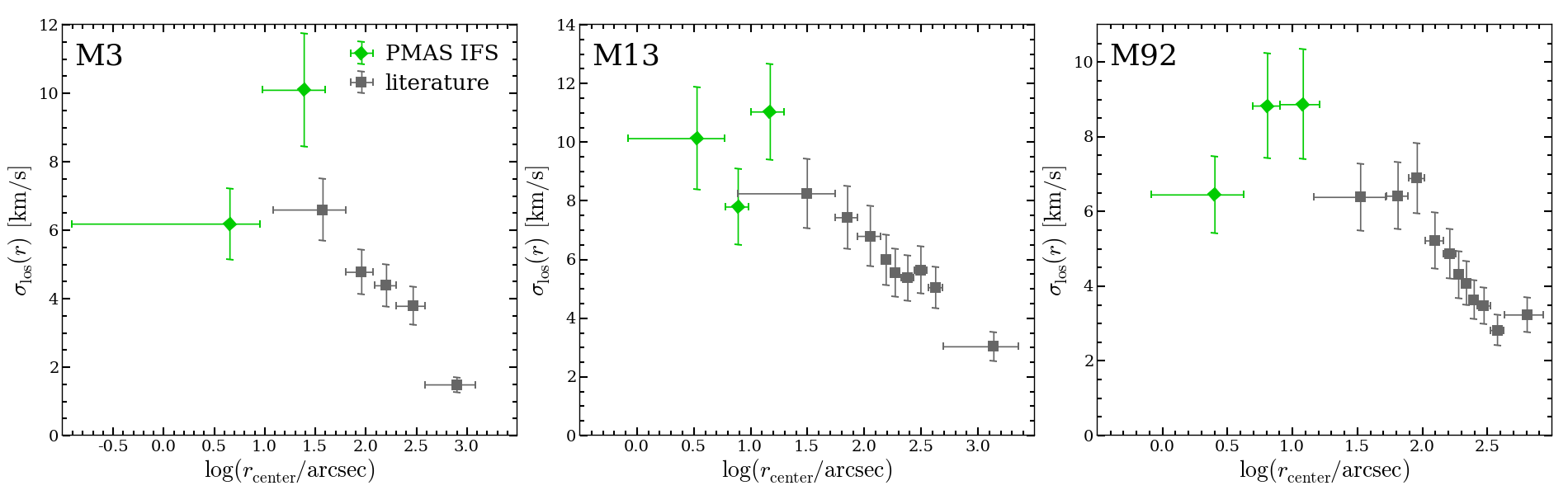}
 \caption{The velocity dispersion as a function of projected distance to the cluster centre for \object{M3}, \object{M13}, and \object{M92}. The PMAS data and the literature data were analysed separately in the sense that no bin included data from both sets. Note that the horizontal bars do not represent errorbars but indicate the range in radii covered by the individual bins.}
 \label{fig:dynamics:binned_dispersion}
\end{figure*}

A disadvantage of the maximum likelihood approach is that it is prone to outliers. In Sect.~\ref{sec:kinematic:rv} we identified several stars in the central cluster regions as potential high-velocity stars, deviating from the mean cluster velocity by more than $3\sigma_\mathrm{0}$, where a literature value for the central dispersion was used. As our PMAS data significantly enhances the kinematic information available near the centre, we re-evaluated the significance of the outliers by using a local estimate of the velocity dispersion, obtained from the $25$ nearest stars in distance to the cluster centre. Only for the candidate in \object{M92} the offset to the mean velocity was larger than $3\times$ this value, while it was smaller for the other candidates. For this reason, we decided to include these stars in the analysis and will discuss their influence separately when comparing them to models in Sect.~\ref{sec:dynamics:jeans}.

We obtained a velocity dispersion profile by binning the data radially, where the bin sizes were chosen so that each bin again contained at least $25$ stars. Figure~\ref{fig:dynamics:binned_dispersion} gives an overview of the profiles obtained this way. Note that the bins were created separately for the PMAS data and the literature data to facilitate a comparison between the two. In general, the agreement is good. Only in \object{M3}, the outer PMAS bin seems to be indicate a significantly higher dispersion than the innermost literature bin.

None of the profiles in Fig.~\ref{fig:dynamics:binned_dispersion} shows a central cusp that could be interpreted as evidence for an intermediate-mass black hole. Instead, the velocity dispersion seems to decrease towards the centre of \object{M3} and \object{M92}. To check whether this decrease is significant, we modelled the dispersion within the core radius determined in Sect.~\ref{sec:photometry:sbprofiles} (cf. Table~\ref{tab:photometry:nuker_fits}) as a simple power-law and found the range of exponents consistent with the data (using the approach outlined in Appendix~\ref{app:data_model_comparison}). We found the trend to be significant in both clusters.

The rise in the velocity dispersion profile that is visible in the very outskirts of \object{M92} was interpreted by \citet{2007AJ....133.1041D} as a consequence of stars escaping from the cluster. No such rise is observed for the other two clusters. \citet{2005ApJS..161..304M} derive tidal radii of $r_\mathrm{t}=178\ \mathrm{pc}$ (M3), $74\ \mathrm{pc}$ (M13), and $59\ \mathrm{pc}$ (M92), respectively. For the distances of the clusters, those values correspond to projected radii of $60\arcmin$, $36\arcmin$ and $23\arcmin$. A comparison with Fig.~\ref{fig:dynamics:binned_dispersion} shows that in \object{M92}, the rise in the velocity dispersion is observed at radii $\gtrsim 1/3 r_\mathrm{t}$. Neither in \object{M3} nor \object{M13} do we have sufficient kinematical data at such large radii to look for a similar effect.

\subsection{Jeans modelling}
\label{sec:dynamics:jeans}

Jeans modelling is a widely used technique to investigate the dynamics of gravitationally bound stellar systems. It is discussed in detail in \citet{2008gady.book.....B}. A good overview is also given in \citet{2010ApJ...710.1063V}. The idea behind Jeans modelling is to estimate the stellar gravitational potential of a cluster from the measured surface brightness profiles via an assumption on the mass-to-light ratio $\Upsilon$. The gravitational potential of a putative IMBH might be added. For a given potential that is spherically symmetric, the Jeans equations yields a prediction of the second order velocity moments of the distribution function of the stars inside the cluster, $\overline{v_\mathrm{r}^2}$, $\overline{v_\mathrm{\theta}^2}$, and $\overline{v_\mathrm{\phi}^2}$. For a comparison with our observations, these moments must be projected into the line of sight. We further require an estimate of the anisotropy in the system to obtain the second order velocity moment along the line of sight, $\overline{v_\mathrm{los}^2}$. The anisotropy is often parametrized via
\begin{equation}
 \beta = 1 - \frac{\overline {v_\mathrm{\theta}^2} + \overline {v_\mathrm{\phi}^2}}{2\overline {v_\mathrm{r}^2}}\,.
 \label{eq:anisotropy}
\end{equation}
A system with preferentially tangential orbits has $\beta<0$, a system with $\beta>0$ has preferentially radial orbits. For an isotropic system, $\overline {v_\mathrm{\phi}^2}=\overline {v_\mathrm{\theta}^2}=\overline {v_\mathrm{r}^2}$.

The second order velocity central moment is the squared sum of the line of sight components of the velocity dispersion and the rotational velocity. The analysis in Sect.~\ref{sec:dynamics:rotation} showed that the rotational component in the clusters is small. For simplicity, we will therefore refer to the projected second order velocity moment as the velocity dispersion and denote it by $\sigma_\mathrm{los}$ in the remainder of the paper.
 
We calculated dynamical models for each of the clusters in our sample. The procedure we followed for every cluster can be summarized as follows. The surface brightness profile of the cluster was parametrized as an extended Nuker profile, cf. Sect.~\ref{sec:photometry:sbprofiles}. Furthermore, we made an assumption about the anisotropy profile $\beta(r)$ in the cluster, starting from an isotropic profile, i.e. $\beta=0$, and considering anisotropy only in \object{M92} where the comparison between the isotropic profile and the data suggested it. For a given mass-to-light ratio $\Upsilon$ and a given black hole mass $M_\mathrm{BH}$ we then determined the radial velocity dispersion profile $\sigma_\mathrm{los}(r)$ as a function of projected distance to the cluster centre and projected into the line of sight. This profile was finally compared to our kinematical data. The code to perform the Jeans modelling was kindly provided by R. van der Marel (priv. comm.). We discuss our results for each cluster in the following.

It should be highlighted that the comparison of our data to the models does not involve binning. When combining the data into radial bins, one has to select a useful bin size. Using too few bins might wash out some characteristics of the data. For example, the upturn in the velocity dispersion at the largest radii in \object{M92} disappears when a size of 50 stars per bin is chosen. On the other hand, too many bins might introduce spurious features in the radial profile due to shot noise resulting from a low number of stars per bin. Therefore we decided to follow the approach by \citet{2002AJ....124.3270G} that does not rely on binning and is outlined in Appendix~\ref{app:data_model_comparison}. The velocity dispersion profiles shown in Fig.~\ref{fig:dynamics:binned_dispersion} will only be used for illustration purposes.

\subsubsection{M3}

\begin{figure*}
 \centering\includegraphics[width=1.\textwidth]{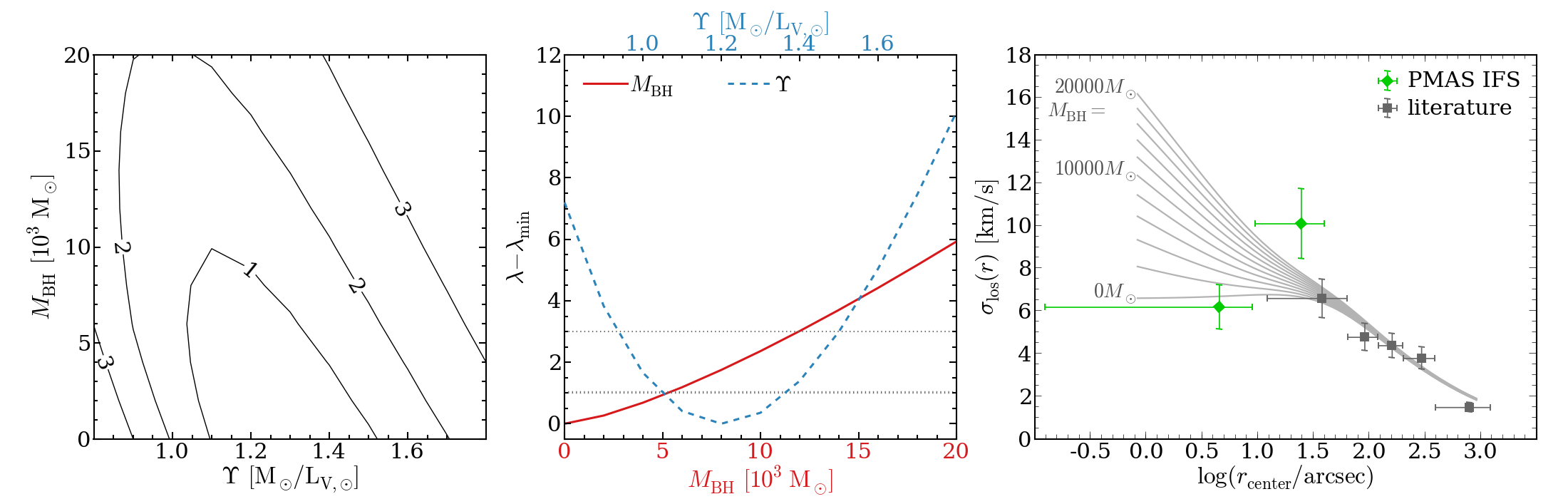}
 \caption{Results of the Jeans modelling of \object{M3}. Models were calculated over a grid of constant mass-to-light ratios $\Upsilon$ and a black-hole masses $M_\mathrm{BH}$  \textit{Left:} Distribution of likelihoods in the $M_\mathrm{BH}$-$\Upsilon$ plane. Contours trace the 1, 2 and 3$\sigma$ confidence intervals. \textit{Centre:} One-dimensional likelihoods as a function of $M_\mathrm{BH}$ (solid line) and $\Upsilon$ (dashed line), respectively, obtained by marginalizing the two-dimensional distribution over the other parameter. Horizontal dotted lines indicate the 1 and 3$\sigma$ confidence limits. \textit{Right:} Comparison between the data and the velocity dispersion profiles predicted by the models with the most likely mass-to-light ratio and a range of black-hole masses, as indicated in the plot.}
 \label{fig:dynamics:m3_jeans}
\end{figure*}

As input for the Jeans models we used constant mass-to-light ratios in the range $\Upsilon=(M_\mathrm{M3}/M_\sun)/(L_\mathrm{V,t}/L_{\mathrm{V},\sun})=0.8\text{--}1.8$ and included contributions of a central black hole in the mass range $0\text{--}20\,000\ \mathrm{M_\odot}$ in the gravitational potential. For each model curve of $\sigma_\mathrm{los}(r)$ we calculated the likelihood according to Eq.~\ref{eq:lambda_stat}. Estimates and confidence intervals for $M_\mathrm{BH}$ and $\Upsilon$ were obtained by marginalizing the likelihoods over the other quantity. The likelihoods of the individual models on the $M_\mathrm{BH}$-$\Upsilon$ grid and the marginalized likelihoods of $M_\mathrm{BH}$ and $\Upsilon$ are shown in Fig.~\ref{fig:dynamics:m3_jeans}. The most likely model has no black hole and a mass to light ratio of $\Upsilon=1.30$.

Marginalizing over models with identical mass-to-light ratio yields $\Upsilon=1.20\substack{+0.17 \\ -0.16}$, corresponding to a cluster mass of $M_\mathrm{M3}=(2.68\pm0.38)\times10^5\ \mathrm{M_\odot}$. The mass-to-light ratio we measure is higher than the value reported by \citet{2005ApJS..161..304M} for the same cluster ($\Upsilon=0.77 \substack{ +0.35 \\ -0.28}$). The offset is explainable, however, as \citeauthor{2005ApJS..161..304M} obtained a dynamical mass-to-light ratio by scaling the velocity dispersion profile predicted by the Wilson model to a central value of $5.6\mathrm{km\,s^{-1}}$, which is smaller than the value predicted by the model that best fits our data. Our model with $\Upsilon=0.8$ and no black hole predicts a central dispersion of $5.5\mathrm{km\,s^{-1}}$, fully consistent with the one used by \citeauthor{2005ApJS..161..304M}.

Our models do not suggest the presence of an intermediate-mass black hole in the cluster centre. We obtain a $1\sigma$ upper limit of $M_\mathrm{BH}<5\,300\ \mathrm{M_\odot}$ when marginalizing over the models with identical black hole mass. Using a more conservative $3\sigma$ limit increases the black hole mass limit to $12\,000\ \mathrm{M_\odot}$. In the right panel of Fig.~\ref{fig:dynamics:m3_jeans} we show a comparison between the velocity dispersion derived from our data and the predictions of the various models with $\Upsilon=1.20$. The distinction between the models with different black hole masses is almost exclusively based on the velocities of the stars in the innermost data bin. The velocity dispersion that we measure in the centre is in good agreement with the model without IMBH. However, none of the model predictions seems to agree with the measured velocity dispersion at distances $\sim20\arcsec$ to the centre. A possible explanation for this are the potential high-velocity stars, as exclusion of the star with the highest relative velocity brings the data in agreement with the model curves and the $1\sigma$ limit on the black hole mass decreases to $2\,200 M_\odot$. On the other hand, the discrepancy may indicate that some of the assumptions in the analysis are not fulfilled, e.g. that the true (kinematic) centre of the cluster is offset from the one we assumed. Unfortunately, our kinematical data in this cluster is limited and also asymmetric with respect to the cluster centre, so we cannot check for this.

We also investigated how the central slope of the surface brightness profile influences our limits on the black hole mass. The profile that provided the best fit to the photometric data was basically a core profile ($\gamma=0.03$), yet shallow cusps could not be excluded in the analysis of Sect.~\ref{sec:photometry:sbprofiles}. We ran the same grid of models as before, but used the maximum cuspy model ($\gamma=0.54$) that is still consistent with the photometric data of \object{M3}. Again no evidence for a black hole was found, instead we obtained a lower $1\sigma$-limit of $1\,700 M_\odot$.

\subsubsection{M13}

\begin{figure*}
 \centering\includegraphics[width=1.\textwidth]{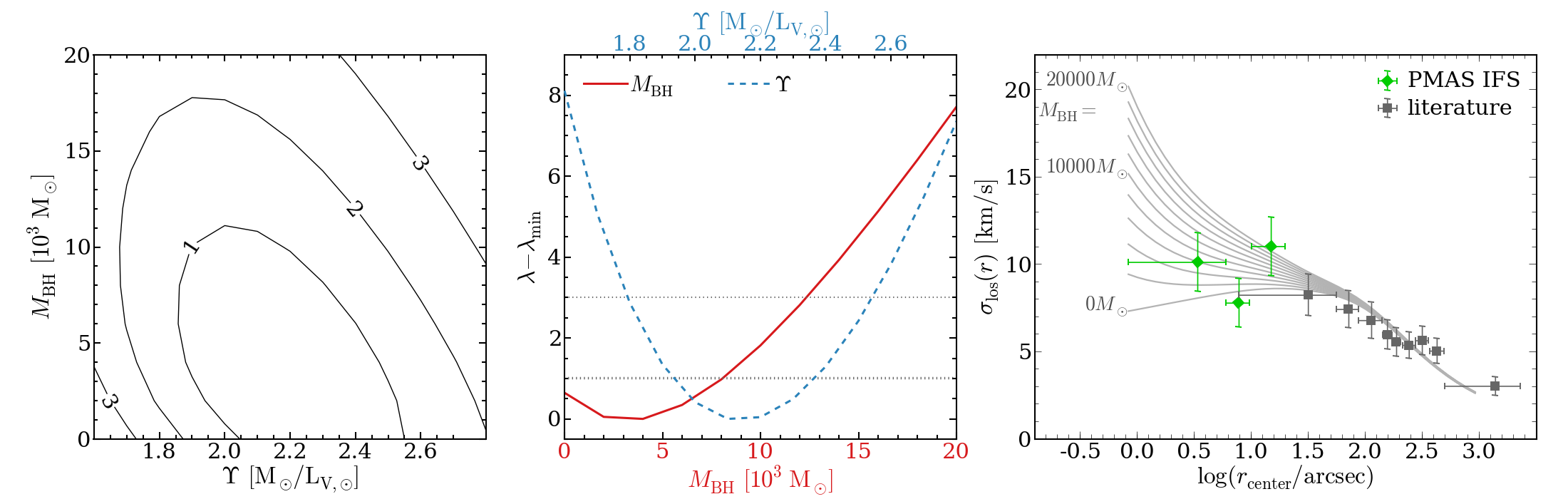}
 \caption{The same as Fig.~\ref{fig:dynamics:m3_jeans} for the globular cluster \object{M13}.}
 \label{fig:dynamics:m13_jeans}
\end{figure*}

The likelihood distribution of the different models calculated for \object{M13} is shown in the left panels of Fig.~\ref{fig:dynamics:m13_jeans}. The most likely model is obtained for a black hole with $M_\mathrm{BH}=2\,000\ \mathrm{M_\odot}$ and a mass-to-light ratio of $\Upsilon=2.20$.

The marginalized mass-to-light ratio estimate of $\Upsilon=2.10\substack{+0.27\\ -0.17}$ that we find is again higher than the value determined by \citet{2005ApJS..161..304M} for the same cluster ($\Upsilon=1.20\substack{+0.54 \\ -0.44}$). The explanation for this discrepancy is the same as for \object{M3}, as \citeauthor{2005ApJS..161..304M} scaled their profile to a central velocity dispersion of $7.1\mathrm{km\,s^{-1}}$ that is below our measurements. Using our estimate of the mass-to-light ratio, we obtain a cluster mass of $M_\mathrm{M13}=(4.02\pm0.42)\times10^5\ \mathrm{M_\odot}$.

The central panel of Fig.~\ref{fig:dynamics:m13_jeans} shows that the difference in likelihoods between including or excluding an intermediate-mass black hole of $2\,000\ \mathrm{M_\odot}$ is not significant. After marginalizing over $\Upsilon$, we obtain an upper limit ($1\sigma$) of $8\,100\ \mathrm{M_\odot}$. A comparison between the velocity dispersion data and the model predictions for different black hole masses in \object{M13} is shown in the right panel of Fig.~\ref{fig:dynamics:m13_jeans}. The measured profile is well described by models with low black hole masses.

We found in Sect.~\ref{sec:photometry:sbprofiles} that the surface brightness profile allows for a range in central slopes, with $\gamma<0.34$. We investigated the influence of the uncertainty in the measured surface brightness profile on our IMBH constraint in the same manner as for \object{M3}. For a model with a core we got only marginal changes in the deduced mass-to-light ratio and upper mass limit for an IMBH. However, if we use the maximum cuspy profile as input to the modelling we find that an IMBH of $M_\mathrm{BH}=4\,000\substack{+4\,600 \\ -3\,200}\ \mathrm{M_\odot}$ yields a better representation of the data at the $2\sigma$ level, compared to the model without a black hole. As a conservative value we therefore adopt an upper mass limit in \object{M13} of $M_\mathrm{BH}<8\,600\ \mathrm{M_\odot}$ ($1\sigma$) or $M_\mathrm{BH}<13\,000\ \mathrm{M_\odot}$ ($3\sigma$).
 
However, we find that the mass range that a putative IMBH can occupy is significantly narrowed if the two potential high velocity stars in this cluster are excluded from the comparison. In that case we find that irrespectively of the assumed central slope of the surface brightness profile, the model that maximizes the likelihood has no black hole, and that every black hole with a mass $>2\,000\ \mathrm{M_\odot}$ is excluded at the $1\sigma$ level.

\subsubsection{M92}

When calculating Jeans models for this cluster we excluded stars with projected distances $>400\arcsec$ to the cluster centre. The reason for this is the observed rise in the velocity dispersion at the largest projected radii (cf. Fig.~\ref{fig:dynamics:binned_dispersion}), probably due to stars escaping from the cluster. The velocities of these stars apparently do not trace the gravitational potential of the cluster any more, but this is the quantity we want to constrain by our models.

\begin{figure*}
 \centering\includegraphics[width=1.\textwidth]{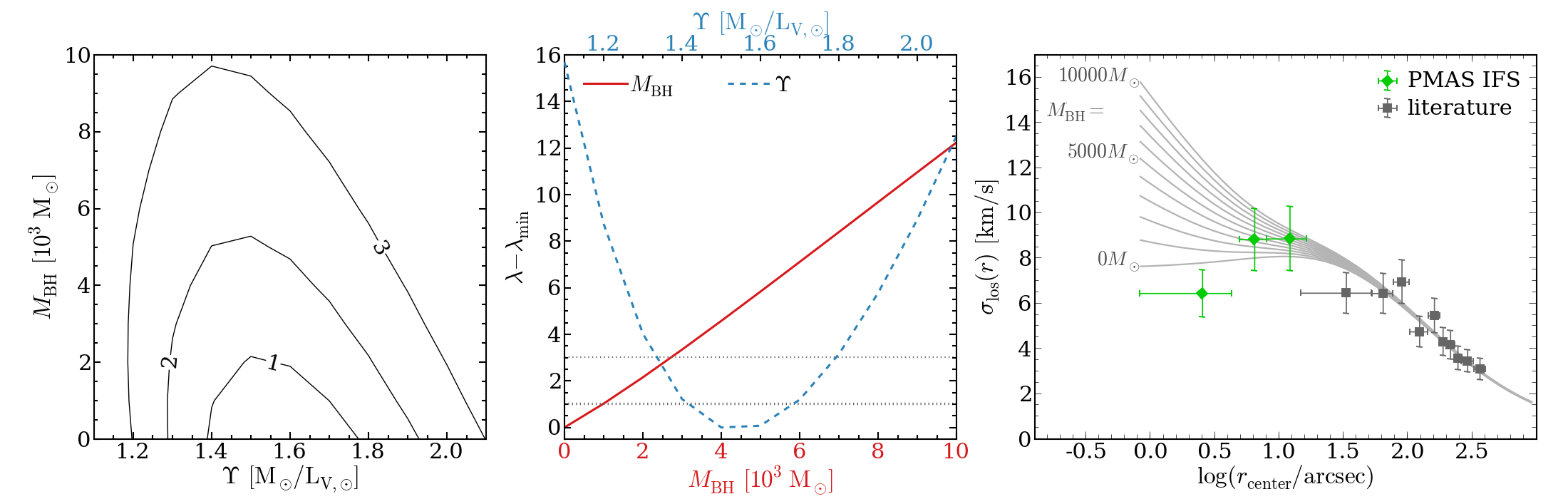}
 \caption{The same as Fig.~\ref{fig:dynamics:m3_jeans} for the globular cluster \object{M92}.}
 \label{fig:dynamics:m92_jeans}
\end{figure*}

We proceeded in analogy to the other two clusters. The distribution of likelihood values is depicted in the left panel of Fig.~\ref{fig:dynamics:m92_jeans}. The model that maximizes the likelihood has no central black hole and a mass-to-light ratio of $\Upsilon=1.60$.

A dynamical mass-to-light ratio of $\Upsilon=1.50\substack{+0.19 \\ -0.09}$ (cf. Fig.~\ref{fig:dynamics:m92_jeans}, centre) is obtained by marginalizing over the black-hole mass. The value is again higher than the value obtained by \citet{2005ApJS..161..304M} for the same cluster, $\Upsilon=0.78\substack{+0.38 \\ -0.27}$, (for the same reason mentioned for \object{M3} and \object{M13}), while it is slightly below the one obtained by \citet{2012A&A...539A..65Z} for a King model, $\Upsilon=1.83$. For our measurement we obtain a cluster mass of $M_\mathrm{M92}=(2.30\pm0.29)\times10^5\ \mathrm{M_\odot}$.

In the right panel of Fig.~\ref{fig:dynamics:m92_jeans} we show the velocity dispersion profiles predicted for models with $\Upsilon=1.50$ and for a range in $M_\mathrm{BH}$ and compare them to the data. As can be seen from the central panel of Fig.~\ref{fig:dynamics:m92_jeans}, our modelling does not favour a massive black hole in the cluster centre, yielding a $1\sigma$ upper mass limit of $980\ \mathrm{M_\odot}$. The $3\sigma$ limit is $2\,700\ \mathrm{M_\odot}$. Our PMAS dataset contains $\sim50$ stars in the region where the models for different black hole masses diverge. Thanks to the more favourable seeing conditions the average S/N of the deblended spectra is also higher than in the other two clusters. This allows us to put a more stringent constraint to the mass of a possible IMBH in the cluster. The overall velocity dispersion profile is well reproduced by the model without a central black hole. There is a slight tendency that even that model overpredicts the central velocity dispersions.

In Sect.~\ref{sec:photometry:sbprofiles} we found that the measured surface brightness profile was best represented using a core profile but that a profile with a central surface brightness cusp $\gamma<0.43$ was also consistent with the data. By computing the same suite of models, but using a cuspy profile with $\gamma=0.43$ as input, we obtained a consistent estimate of the mass-to-light ratio, $\Upsilon=1.50\substack{+0.09 \\ -0.17}$ and our $1\sigma$ limit for the mass of a black hole decreased only slightly ($820\ \mathrm{M_\odot}$).

For this cluster, we also compared our data to the self-consistent models calculated by \citet{2007MNRAS.381..103M}. These are based on King profiles with different black hole masses added. We obtained model predictions for a central potential of $W_\mathrm{0}=7.7$ \citep[as determined by][]{2013ApJ...774..151M} and mass ratios of black hole to cluster mass ranging from $0$ to $0.9\%$. For each profile, we determined the scale radius by fitting the predicted surface brightness profile to the data within the half-light radius obtained by \citet{2013ApJ...774..151M}. Then, we compared the predicted velocity dispersion profiles to our data via the same maximum likelihood approach used for the Jeans models. We find that models with IMBH masses $<0.6\%$ of the cluster mass provide an acceptable fit to our data. Using our estimate of the cluster mass, this limit corresponds to $\sim1\,400\ \mathrm{M_\odot}$, in good agreement with our limit obtained from the Jeans models.

\section{Discussion}
\label{sec:discussion}

We do not find strong evidence for the presence of intermediate-mass black holes in the clusters. \object{M13} seems to be the most ambiguous case, as our upper limit is least stringent in this cluster and the presence of an IMBH could even be preferred if we adopt a surface brightness profile with the steepest cusp consistent with the photometric data. However, the ambiguity can largely be attributed to the presence of two potential high velocity stars, as excluding them results in an upper limit $\sim2\,000\ \mathrm{M_\odot}$ regardless of the assumed surface brightness profile. However, our current dataset does not allow us to exclude these stars on firm statistical grounds. \object{M13} may be considered the best candidate to search for such a black hole, using data with a higher resolution compared our PMAS cubes.

We now compare our measurements in Table~\ref{tab:discussion:imbhs_in_gcs} to other searches for intermediate-mass black holes in globular clusters. Only studies based on the internal cluster dynamics are considered. It should also be mentioned that Table~\ref{tab:discussion:imbhs_in_gcs} includes clusters for which conflicting results were obtained. Therefore care must be taken when interpreting the data. The majority of the work done so far focussed on the most massive clusters in the Milky Way (or \object{M31} in the case of \object{G1}). While \object{M3}, \object{M13}, and \object{M92} still belong to the more massive part of the overall globular cluster population in the Milky Way, they cover the lower end of the mass range studied so far. Compared to the limits on black hole masses found for more massive clusters, the upper limit we find for \object{M92} is very stringent. To our knowledge, it is the lowest one determined so far for a massive globular cluster from the kinematics of resolved stars. This is quite remarkable given that our observations were carried out from the ground under modest seeing conditions with a 4m-class telescope. It shows the power of our crowded field 3D spectroscopy approach.

The spherical Jeans modelling approach employed by us to interpret the kinematics of our sample clusters does involve some simplifications. We now discuss the most important of those and check whether they are likely to affect our results significantly.

A constant mass-to-light ratio $\Upsilon$ was assumed throughout our dynamical modelling. This is a simplification because the dynamical evolution of the cluster will lead to mass segregation. A rise of $\Upsilon$ in the outskirts is expected because of the increased abundance of low-mass stars. At the smallest radii, the situation might be more complex. The accumulation of heavy stellar remnants has been predicted to cause a central spike in the $\Upsilon$ profile. \citet{1995AJ....109..209G} found such a behaviour only in very concentrated clusters, where the mass-to-light ratio had a minimum at distances around $1\arcmin$ towards the centre. In less concentrated clusters, $\Upsilon$ decreased monotonically with radius. A central increase in $\Upsilon$ would also increase the central velocity dispersion predicted by the model. In the cases of \object{M3} and \object{M92}, such an increase would reduce the agreement with our measurements. So our data do not suggest that $\Upsilon$ increases towards the centre of either \object{M3} or \object{M92}.

\begin{table*}
 \centering
 \caption[]{\label{tab:discussion:imbhs_in_gcs}Intermediate-mass black holes in globular clusters. Constraints from kinematical studies.}
 \begin{tabular}{ c c c c c c c c l }
\hline\hline
Name &
method &
$M_\mathrm{GC}$ &
$\epsilon_\mathrm{M_{GC}}$ &
$\sigma_\mathrm{0}$ &
$\epsilon_{\sigma_\mathrm{0}}$ &
$M_\mathrm{IMBH}$ &
$\epsilon_\mathrm{M_{IMBH}}$ &
ref. \\
&
 &
$\mathrm{M_\odot}$ &
$\mathrm{M_\odot}$ &
$\mathrm{km}\,\mathrm{s}^{-1}$ &
$\mathrm{km}\,\mathrm{s}^{-1}$ &
$\mathrm{M_\odot}$ &
$\mathrm{M_\odot}$ &
\\ \hline
 47\,Tuc & p.m. & $1.1 \times 10^6$ & $1.0 \times 10^5$ & $11.5$ & $2.3$ & $<1.5 \times 10^3$ &  & 1\\
 NGC\,1851 & l.b. & $3.7 \times 10^5$ & $2.6 \times 10^4$ & $9.3$ & $0.5$ & $<2.0 \times 10^3$ &  & 2\\
 NGC\,1904 & l.b. & $1.4 \times 10^5$ & $1.0 \times 10^4$ & $8.0$ & $0.5$ & $3.0 \times 10^3$ & $1.0 \times 10^3$ & 2\\
 NGC\,2808 & l.b. & $8.5 \times 10^5$ & $1.0 \times 10^5$ & $13.4$ & $2.6$ & $<6.0 \times 10^3$ &  & 3\\
 $\omega\,Cen$ & p.m. & $2.5 \times 10^6$ & $2.3 \times 10^5$ & $16.0$ & $3.2$ & $<1.2 \times 10^4$ &  & 4\\
 $\omega\,Cen$ & l.b. & $2.5 \times 10^6$ & $2.3 \times 10^5$ & $16.0$ & $3.2$ & $4.7 \times 10^4$ & $1.0 \times 10^4$ & 5\\
 NGC\,5286 & l.b. & $4.4 \times 10^5$ & $1.8 \times 10^4$ & $10.0$ & $2.0$ & $1.5 \times 10^3$ & $1.0 \times 10^3$ & 6\\
 NGC\,5694 & l.b. & $2.6 \times 10^5$ & $3.0 \times 10^4$ & $8.8$ & $0.6$ & $<8.0 \times 10^3$ &  & 2\\
 NGC\,5824 & l.b. & $4.5 \times 10^5$ & $3.1 \times 10^4$ & $11.2$ & $0.4$ & $<6.0 \times 10^3$ &  & 2\\
 NGC\,6093 & l.b. & $3.4 \times 10^5$ & $1.6 \times 10^4$ & $9.3$ & $0.3$ & $<8.0 \times 10^2$ &  & 2\\
 NGC\,6266 & p.m. & $8.2 \times 10^5$ & $1.7 \times 10^4$ & $13.7$ & $1.1$ & $<4.0 \times 10^3$ &  & 7\\
 NGC\,6266 & l.b. & $9.3 \times 10^5$ & $2.1 \times 10^4$ & $15.5$ & $0.5$ & $2.0 \times 10^3$ & $1.0 \times 10^3$ & 2\\
 NGC\,6388 & l.b. & $1.1 \times 10^6$ & $1.7 \times 10^5$ & $18.9$ & $0.8$ & $1.7 \times 10^4$ & $9.0 \times 10^3$ & 8\\
 NGC\,6388 & r.v. & $1.1 \times 10^6$ & $1.7 \times 10^5$ & $13.0$ & $2.0$ & $<2.0 \times 10^3$ &  & 9\\
 M15 & r.v., p.m. & $6.0 \times 10^5$ & $4.0 \times 10^4$ & $12.0$ & $2.0$ & $<3.4 \times 10^3$ &  & 10\\
 G1 & l.b. & $1.1 \times 10^7$ & $4.4 \times 10^6$ & $25.1$ & $1.7$ & $1.8 \times 10^4$ & $5.0 \times 10^3$ & 11\\
 M3 & r.v. & $2.7 \times 10^5$ & $3.8 \times 10^4$ & $7.9$ & $0.7$ & $<5.3 \times 10^3$ &  & This study \\
 M13 & r.v. & $4.0 \times 10^5$ & $5.2 \times 10^4$ & $9.3$ & $0.8$ & $<8.1 \times 10^3$ &  & This study \\
 M92 & r.v. & $2.3 \times 10^5$ & $2.9 \times 10^4$ & $8.1$ & $0.7$ & $<9.8 \times 10^2$ &  & This study \\
\hline
\end{tabular}
\tablebib{
(1) \citet{2006ApJS..166..249M}; (2) \citet{2013A&A...552A..49L}; (3) \citet{2012A&A...542A.129L}; (4) \citet{2010ApJ...710.1063V}; (5) \citet{2010ApJ...719L..60N}; (6) \citet{2013A&A...554A..63F}; (7) \citet{2012ApJ...745..175M}; (8) \citet{2011A&A...533A..36L}; (9) \citet{2013ApJ...769..107L}; (10) \citet{2006ApJ...641..852V}; (11) \citet{2005ApJ...634.1093G}
}

 \tablefoot{Column 2 contains the method(s) used to infer stellar dynamics: proper motions (p.m.), radial velocities (r.v.), or line broadening (l.b.). The upper limits on IMBH masses are $1\sigma$ limits.}
\end{table*}

It is not fully understood how the inclusion of a dedicated mass-to-light ratio profile influences the search for an IMBH. While \citet{2002AJ....124.3270G,2003AJ....125..376G} found that in \object{M15} the need for an IMBH is diminished when including a model-predicted profile, \citet{2012A&A...542A.129L} observed the opposite trend in \object{NGC~2808}. Note that in the latter case the largest changes occurred at radii $\gtrsim10\arcsec$ where the predicted velocity dispersion profile underestimated their data. So their higher IMBH mass limit could also be a consequence of this mismatch.

We do not expect that the assumption of a constant mass-to-light ratio profile has led us to significantly underestimate the upper limits on the black hole masses that are large determined by the innermost datapoints. Minor discrepancies between the data and our models at larger radii, such as observed in \object{M92} at distances $\sim10\arcsec-100\arcsec$, might be hinting towards a more complex M/L profile. It is possible to obtain such a profile for example via Fokker-Planck models \citep[e.g.][for \object{M15}]{1997ApJ...481..267D,2003ApJ...585..598D} or by N-body simulations. However, this is beyond the scope of this work. On the other hand, if all three components of the velocity dispersion are known, it is also possible to constrain the profile of the mass-to-light ratio directly from observations. Such has been done by \citet{2006ApJ...641..852V} for \object{M15} or by \citet{2006A&A...445..513V} for \object{$\omega$~Centauri}. The combination of our data with proper motions could be very promising in this respect.

We observed only one component of the velocity dispersion, namely the one along the line of sight. For this reason we cannot infer whether the velocity distribution is anisotropic in any of our target clusters. Anisotropy has been identified as a significant nuisance in the search for massive black holes. For example, a radially biased velocity distribution can mimic the presence of a black hole \citep{2008gady.book.....B}. On the other hand, any initial anisotropy in a stellar system will decrease with time due to relaxation processes. The high stellar densities in the cluster centres imply relaxation timescales that are only a small fraction of the lifetime of a cluster \citep[e.g.][]{1988A&A...191..215M}. While evidence for anisotropy has been detected in \object{M3} and \object{M13}, it is restricted to larger radii, so we do not expect this to strongly affect our analysis of the central kinematics.
However, the presence of a massive black hole can lead to a tangential bias because stars on high-eccentricity orbits are ejected by the black hole \citep{2003ApJ...583...92G}. In  \object{M92} there is a tendency that the model prediction overestimates the velocity dispersion in the central region. Such a trend might be caused by tangential anisotropy in the centre. To test whether our data supports such a scenario, we ran anisotropic Jeans models for \object{M92} with $\beta(r)$ parametrized as
\begin{equation}
 \beta(r) = \beta_\mathrm{0} + (\beta_{\infty}-\beta_\mathrm{0})\frac{r^2}{r^2+a^2}\,.
 \label{eq:beta_function}
\end{equation}
The anisotropy radius $a$ was fixed to the core radius ($\sim$10\arcsec, cf. Table~\ref{tab:photometry:nuker_fits}) and $\beta_{\infty}$ was set to zero because the cluster is known to be isotropic at larger radii \citep{1976AJ.....81..975C}. No black hole was included, the mass-to-light ratio was fixed to the value that maximized the likelihood in the isotropic case, and $\beta_\mathrm{0}$ was varied between $-3$ and $0$. We find a confidence interval of $\beta_\mathrm{0}=-0.6\substack{-0.8 \\+0.7}$. Thus, while our data may suggest tangential anisotropy in the core, we do not detect it significantly.

A valid dynamical modelling of the central regions depends critically on an accurate determination of the centre of the cluster. As mentioned earlier, we adopted the photometric cluster centres determined by \citet{2010AJ....140.1830G} using two different methods. A correct localization of the centre is more difficult in clusters with a shallow central density gradient. In this respect, the large core radii found in \object{M3} and \object{M13} might be problematic. Yet \citeauthor{2010AJ....140.1830G} give small uncertainties for all clusters that we studied ($0\farcs2$ for \object{M3}, $0\farcs1$ for \object{M13} and $0\farcs3$ for \object{M92}). Our analysis is based on the assumption that the photometric centre of each cluster coincides with the kinematic centre. The peculiarity of distinct kinematical and photometric centres has been discussed for $\omega$ Centauri \citep{2010ApJ...719L..60N} which seems far more complex than our clusters. It might be an explanation for the deviation we observe between our data and the models in \object{M3}. However, our data do not cover a large enough region of the central region to test this.

\section{Conclusions}
\label{sec:conclusions}

We introduce crowded field 3D spectroscopy as a powerful tool to study the central kinematics of Galactic globular clusters. The combination of PMAS integral field data with PSF fitting techniques allows us to gather large samples of stellar spectra in the highly crowded centres of the three clusters \object{M3}, \object{M13}, and \object{M92}. In our seeing limited datasets, individual stars are resolvable down to the main sequence turn-off of each cluster. As expected, the S/N of the deblended spectra decreases as the stars get fainter, yet reliable radial velocity measurements are still possible for stars below the horizontal branch. The combination of our new measurements with literature data allows us to gather kinematic information at all cluster radii. All our catalogues of radial velocity measurements are made available in the online version of this paper.

Our PMAS stars constitute by far the largest spectroscopic samples obtained in the central $10$--$20\arcsec$ of any of the three clusters to date. They allow us for the first time to put meaningful constraints on the presence of intermediate-mass black holes in these clusters. Using spherical Jeans models, we find that a massive black hole is required in none of them to reach agreement between the Jeans models and our data. Our analysis thus puts stringent upper limits on the mass of any such object in the investigated clusters. In \object{M92} we can rule out an intermediate-mass black hole more massive than $2\,700\ \mathrm{M_\odot}$ at the $3\sigma$ level.

Our $3\sigma$ upper limits for \object{M3} ($12\,000\ \mathrm{M_\odot}$) and \object{M13} ($13\,000\ \mathrm{M_\odot}$) are more relaxed. \object{M13} remains the least conclusive object we investigated. In contrast to the other two clusters, the most likely model for this cluster does contain a black hole (of $2\,000\ \mathrm{M_\odot}$). However, the improvement over the model without a black hole is insignificant. In this respect, it is unfortunate that the kinematical data in the centre of \object{M13} suffered from non-optimal observing conditions. In summary, we consider \object{M13} the most promising candidate to potentially host an IMBH, still below our sensitivity limit but possibly detectable with better data.

The analysis of the central kinematics in the two clusters \object{M3} and \object{M13} is complicated by the presence of a few potential high velocity stars. Excluding these stars from the analysis would strongly narrow the range of black hole masses allowed by our modelling. Our current dataset is insufficient to discriminate whether these stars represent the wings of the internal velocity dispersion or have been accelerated by other means. Acceleration by an undetected intermediate-mass black hole is unlikely in view of the relatively large distances to the centres and the sizes of the stellar samples \citep{2003ApJ...597L.125D}. As mentioned earlier, another possible scenario involves the acceleration by stellar mass black holes. Heating by stellar mass black holes has also been suggested as an explanation for large core radii \citep{2008MNRAS.386...65M,2012Natur.490...71S}. In this respect, the large core radii of \object{M3} \citep[$1.1\mathrm{pc}$, ][]{2005ApJS..161..304M} and \object{M13} ($1.2\mathrm{pc}$) are intriguing, they may suggest a wealthy black hole population in the centres of the clusters and larger kinematic samples will be able to investigate this.

The search for intermediate-mass black holes in globular clusters has gained a lot of attention over the last years. Many of the recent studies rely on observations at the highest possible spatial resolutions. In the future such observations may be used to address the questions beyond the sensitivity of our dataset. However, the requirement for adaptive optics or multi-epoch HST observations makes them rather expensive in terms of observing time. We demonstrated in this work that meaningful results can be obtained based on less demanding observations. It is also clear that significant further improvements can be expected from applying the technique of crowded field 3D spectroscopy to higher quality data.

\begin{acknowledgements}
 The authors are grateful to Roeland van der Marel for sharing his Jeans modelling code.\newline{}
 S.~K. acknowledges support from the ERASMUS-F project through funding from PT-DESY, grant no. 05A09BAA, and from DFG via SFB 963/1 ``Astrophysical flow instabilities and turbulence'' (project A12). \newline{}
 C.~S. was supported by funds of DFG and Land Brandenburg (SAW funds from WGL), and also by funds of PTDESY-0512BA1. \newline{}
 Part of this work is based on observations made with the NASA/ESA Hubble Space Telescope, and obtained from the Hubble Legacy Archive, which is a collaboration between the Space Telescope Science Institute (STScI/NASA), the Space Telescope European Coordinating Facility (ST-ECF/ESA) and the Canadian Astronomy Data Centre (CADC/NRC/CSA). \newline{}
 Some of the data presented in this paper were obtained from the Mikulski Archive for Space Telescopes (MAST). STScI is operated by the Association of Universities for Research in Astronomy, Inc., under NASA contract NAS5-26555. Support for MAST for non-HST data is provided by the NASA Office of Space Science via grant NNX09AF08G and by other grants and contracts. \newline{}
 This research has made use of the products of the Cosmic-Lab project funded by the European Research Council.
\end{acknowledgements}

\appendix

\section{Data reduction}
\label{app:reduction}

\subsection{Scattered light subtraction}

\begin{figure*}
 \centering\includegraphics[width=17cm]{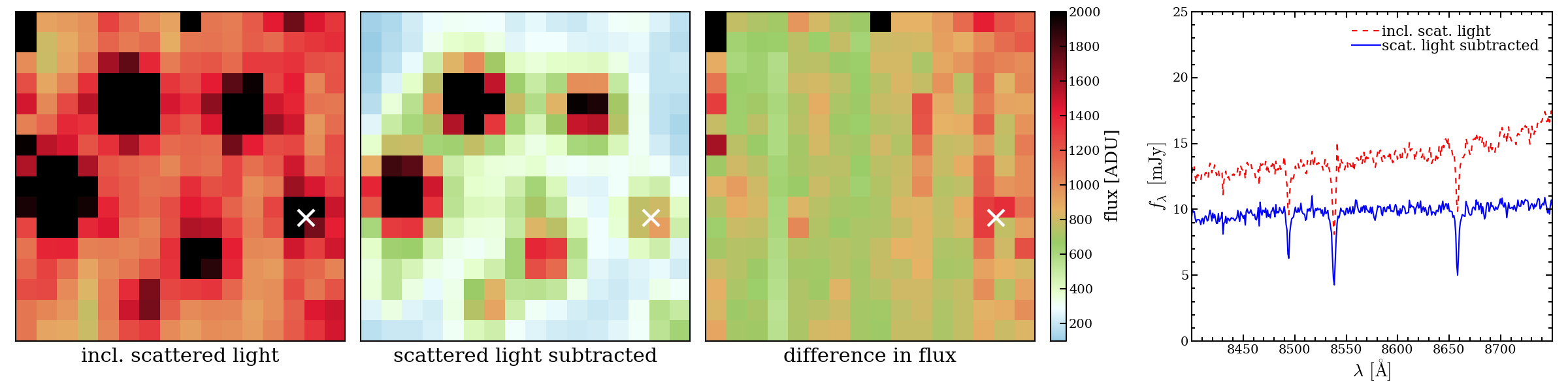}
 \caption{Effect of scattered light in our PMAS data. For a reduced data cube of our central PMAS pointing in M3 the first three panels show, from left to right, a white light image with the scattered light still included, a white light image with the scattered light subtracted, and the difference in flux between the two. All images are displayed on the same intensity scale indicated by the colour bar on the right-hand side of the plot. The right panel compares the spectra extracted for the star marked by a white cross in the PMAS data before and after the scattered light subtraction.}
 \label{fig:dr:scatlight_demo}
\end{figure*}

Besides the traces of the individual science fibres and the bias level, we found an additional light component in each of our exposures, scattered all over the CCD. The origin of this straylight component is unknown. It may be light scattered in the spectrograph or result from problems with the CCD controller \citep{2013A&A...549A..87H}. However, it varies in strength and can reach intensities comparable to the science signal. Therefore we had to develop a method to subtract it. Using the trace mask, i.e. the position of each fibre in cross-dispersion direction as a function of wavelength, we first masked out all pixels around the traces. The remaining pixels should then contain only the scattered light. For each column of the CCD, we modelled the scattered light along the cross-dispersion direction by fitting a low order polynomial to the unmasked pixels. To obtain a smooth representation of the scattered light also as a function of wavelength, the sequence of one-dimensional fits was finally smoothed with a Gaussian kernel. This procedure has meanwhile also been implemented into \textsc{p3d} \citep{2012SPIE.8451E..0FS}.

In Fig.~\ref{fig:dr:scatlight_demo} we show the drastic improvement in data quality that we achieved when we accounted for scattered light. The difference image shows that the scattered light was not just a spatially flat component in the reduced datacubes, but strongly influenced the relative amount of light that was extracted for individual fibres. Remember that our analysis relies on determining the PSF in an observation from the datacube and using this information to deblend stellar spectra. Any artefact in the data that changes the relative intensities of spaxels would have a severe influence on the determination of the PSF because the intensities of the individual spaxels are no more governed by the shape of the PSF. Therefore, it is no surprise that the scattered light strongly biases the spectra that are extracted for the individual stars. This can be verified from the two spectra that are shown in the rightmost panel of Fig.~\ref{fig:dr:scatlight_demo} that were extracted for the same star, before and after the scattered light subtraction.

\subsection{Flat fielding}
In order to obtain correct relative spaxel intensities, a high quality fibre-to-fibre flat fielding of the data is essential. A problem of PMAS is that calibration data obtained during the night cannot be used to correct for the different efficiencies of the individual fibres because the calibration lamps do not illuminate the lens array homogeneously. Therefore one has to rely on twilight flats. However, PMAS is mounted on the Cassegrain focus of the telescope, thereby it is strongly affected by flexure. As a consequence, the twilight flats cannot be used to correct the data for the fringing of the CCD because the fringing pattern is time-dependent.

We applied the following flat fielding procedure: Starting from the reduced twilight flat, we divided each fibre signal by the mean of all fibres and fitted the resulting curves with Chebyshev polynomials of order around 10. Each science datacube was then flat fielded using the polynomial fits. This step corrects for the different (wavelength-dependent) transmissivities of the fibres, but it does not correct for the fringing. The fringing pattern was removed using the night-time calibration data: each spectrum in the continuum flat, originally used to trace the spectra across the CCD, was also fitted with a Chebyshev polynomial and afterwards divided by the fit. As a result, we obtained a normalized spectrum for each fibre that still included the fringes. Division of the science data by these spectra finally removed the fringes.

\section{Checking the completeness of the ACS data}
\label{app:catalog_correction}

To cross-check the photometry we searched for data obtained with WFPC2. While probably being less accurate compared to ACS, WFPC2 data do not suffer as strongly from bleeding artefacts. For \object{M13}, we used the available photometry from \citet{2002A&A...391..945P}. Unfortunately, neither \object{M3} nor \object{M92} were observed during this campaign. For those two clusters, however, we obtained raw $V$- and $I$-band photometry from the HST archive and analysed them using \textsc{dolphot} \citep{2000PASP..112.1383D}. The available photometry for \object{M13} covers the $B$- and $V$-bands, but our spectra are in the near infrared and $I$-band magnitudes were used in the source selection process for the deblending. To get $I$-band magnitudes for the \object{M13} stars, we used the isochrones of \citet{2008A&A...482..883M}, available for a wide range of magnitudes in the HST photometric system. We compared the magnitudes in $B$ and $V$ for every star in the HST photometry to those in the isochrone and then assigned it the $I$-band magnitude of the best match in the isochrone.

\begin{figure}
 \resizebox{\hsize}{!}{\includegraphics{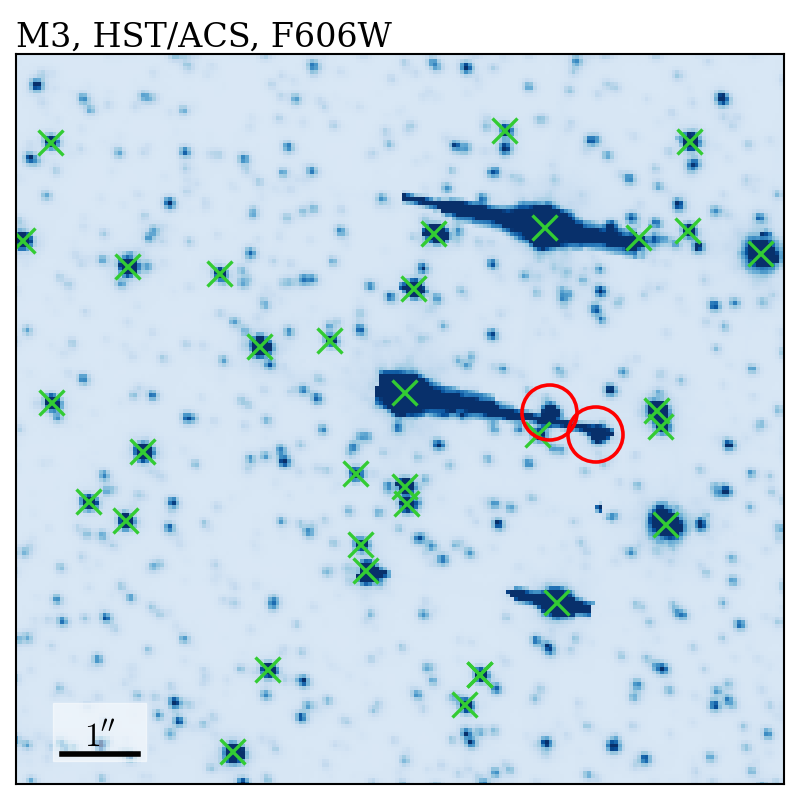}}
 \caption{Illustration of incompleteness in the ACS input catalogue, using the central region of \object{M3}. Green crosses indicate sources included in the ACS catalogue, sources marked by red circles were added after a comparison with archival \object{WFPC2} photometry. For clarity, only stars with $I$-band magnitudes brighter than 17 are highlighted. The cluster centre is located directly on the central bright star.}
 \label{fig:appb:m3_added_stars}
\end{figure}

\begin{figure*}
 \centering \includegraphics[width=17cm]{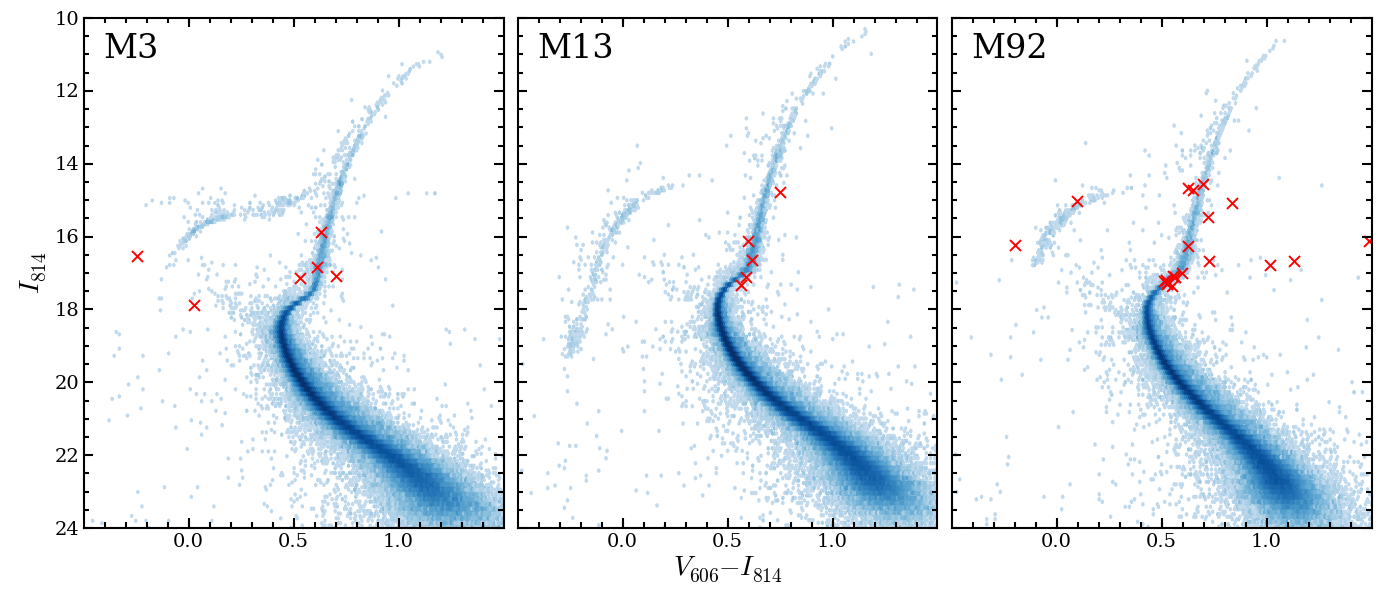}
 \caption{Colour-magnitude diagrams of \object{M3} (left), \object{M13} (centre), and \object{M92} (right). The photometry obtained in the ACS survey of Galactic globular clusters is shown as a blue density plot. Red crosses mark the individual stars we added based on the analysis of archival HST WFPC2 data.} 
 \label{fig:appb:corrected_cmds}
\end{figure*}

As expected, the ACS catalogue is quite complete at the magnitude levels that we are most interested in. However, we find that occasionally giant stars are missing. The reason for this becomes clear in Fig.~\ref{fig:appb:m3_added_stars}, where we illustrate the result of our cross checking procedure for \object{M3}. The image shown in Fig.~\ref{fig:appb:m3_added_stars} is a combination of all data used to create the ACS catalogue. If bright stars were located directly on the bleeding artefacts caused by even brighter ones, they might have been missed. To avoid this, a single short exposure was taken as part of the ACS survey for every cluster observed. However, ACS has a gap of some arcseconds width in between the two chips and no short exposure data exist in that area. The stars we additionally detect using the second catalogue lie preferentially in the same area and are highlighted in the colour magnitude diagrams in Fig.~\ref{fig:appb:corrected_cmds}. We note that some of the added stars are offset from the cluster population, especially for \object{M92}. We verified that those are not spurious detections. However, their measurements might have been affected by the presence of nearby stars that are significantly brighter.

\section{Processing of literature data}
\label{app:literature}

For each star in each catalogue listed in Table~\ref{tab:kinematic:literature_studies}, we determined the weighted mean velocity and its uncertainty from the available data. This combination process also offers a convenient way to detect stars with variable radial velocities. For the scatter of the individual measurements to be consistent with the uncertainties, their $\chi^2$ value should be comparable to the number of degrees of freedom $N_\mathrm{dof}$, i.e., the number of measurements minus one. For any combination of $\chi^2$ and $N_\mathrm{dof}$, the probability of consistency can be calculated from the probability density function of the $\chi^2$ distribution. We identified as ``RV variable'' all stars with probabilities less than $1\%$ and absolute offsets $\geq2\,\mathrm{km\,s^{-1}}$.

\begin{figure*}
 \centering
 \includegraphics[width=17cm]{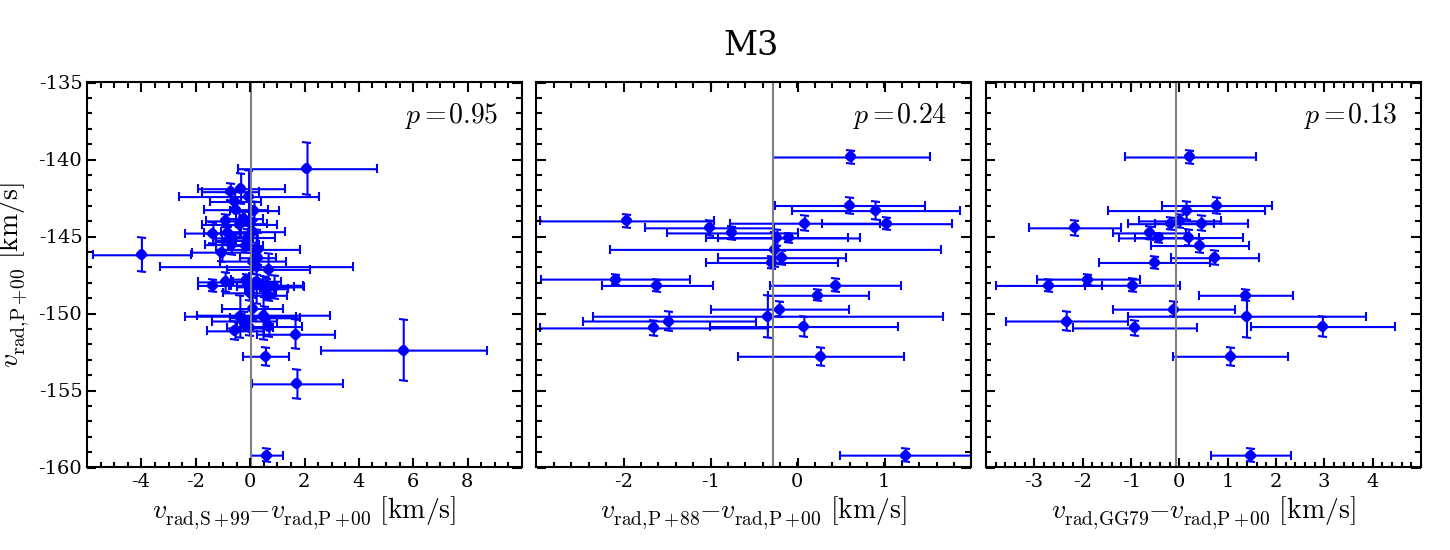}
 
 \includegraphics[width=17cm]{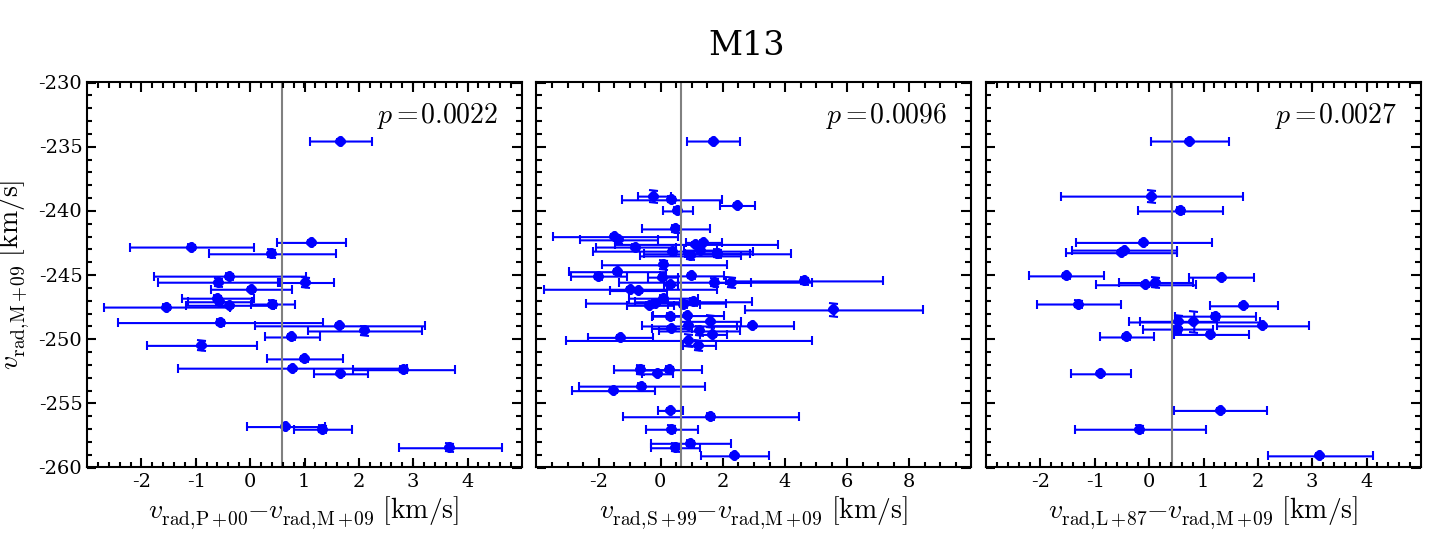}

 \includegraphics[width=17cm]{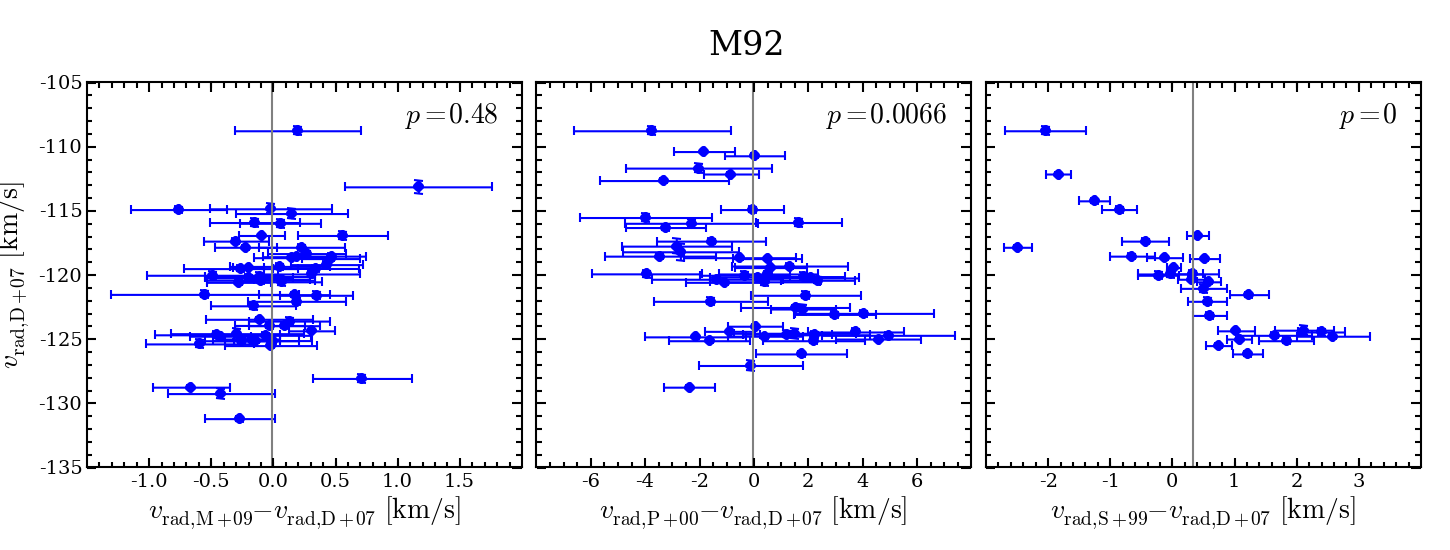}
 \caption{Comparison of the different literature datasets in \object{M3} (top), \object{M13} (centre), and \object{M92} (bottom). The different panels show the offsets in the measured radial velocities between the various studies and the reference study for stars present in both samples. The probability that the scatter of the individual offsets is consistent with the provided uncertainties is given in the upper right corner of each panel. The abbreviations used to denote the individual studies are given in Table~\ref{tab:kinematic:literature_studies}.}
 \label{fig:appc:literature_comparison}
\end{figure*}

Any meaningful combination of the velocities relies on the assumption that the individual studies yield consistent results. We checked that this is the case by defining a reference dataset for each cluster and identifying the subset of stars that was also observed by one or more of the other studies. For this subset, which contained $\gtrsim20$ stars for any given combination of two studies, we computed the differences in the measured radial velocities. In the case of consistent results, the differences should scatter around zero. If the mean of the distribution deviates from zero, this indicates different assumed systemic cluster velocities, a minor inconsistency that can be easily corrected for. We show the results of this comparison in Fig.~\ref{fig:appc:literature_comparison}. In most of the cases, the measured offsets were small ($\lesssim 1 km\,s^{-1}$) and no systematic trends were observed. A notable exception are the data of \citet{1999PASP..111.1233S} in \object{M92}. As already mentioned by \citet{2007AJ....133.1041D}, the velocities measured are biased towards the cluster mean, a trend likely caused by the presence of interstellar absorption lines in the spectra. We observed the same trend with respect to both comparison catalogues and therefore decided to omit this dataset. However, neither in \object{M13} nor in \object{M3} did we observe a similar trend, likely because the spectra covered a different wavelength range. Therefore, we kept the data of \citeauthor{1999PASP..111.1233S} for those two clusters.

After checking the measured velocities for their reliability, we used the comparison shown in Fig.~\ref{fig:appc:literature_comparison} to further check whether the \emph{uncertainties} provided in the different studies are reliable. In that case, the scatter of the individual offsets should again be consistent with the measurement errors, i.e., the $\chi^2$ value should be of the order of the number of degrees of freedom, in this case the number of stellar pairs minus one. Radial velocity variables will bias this comparison. Therefore we iteratively cleaned the comparison samples from such stars. For each combination plotted in Fig.~\ref{fig:appc:literature_comparison} the probability that the scatter is consistent with the uncertainties is provided. The probabilities are sometimes quite small, indicating that the uncertainties may be underestimated. In such cases we applied a constant correction factor to the uncertainties that yielded a $\chi^2$ equal to the number of degrees of freedom. In doing so, we started from the data of \citet{2007AJ....133.1041D} in \object{M92} as it contains by far the largest sample of stars and uncertainties were determined very carefully. We then determined the correction factors successively for all other studies. The factors that we obtained were small ($\sim1.2$--$1.5$).

In the final step, we combined the results of the individual samples and again flagged all stars that showed evidence for radial velocity variations, using the same criteria as before. This yielded the final set of stellar velocities that we used in our subsequent analysis.

\section{Comparing data to models}
\label{app:data_model_comparison}

When a model prediction of ($v_\mathrm{sys}$, $\sigma_\mathrm{los}$) is available as a function of radius, as we obtain from our Jeans models, it is possible to investigate the agreement between the model and the kinematical data without binning the latter. To do so, the likelihood of each model given the data is computed under the assumption that each measurement is drawn from the probability distribution
\begin{equation}
 f(v_{\mathrm{rad},i}) = \frac{1}{\sqrt{2\pi(\sigma_\mathrm{los}^2+\epsilon_{\mathrm{v_{rad}},i}^2)}}\,\exp\left\{-\frac{(v_{\mathrm{rad},i}-v_\mathrm{sys})^2}{2(\sigma_\mathrm{los}^2+\epsilon_{\mathrm{v_{rad}},i}^2)}\right\}\,.
 \label{eq:vel_distribution}
\end{equation}
Recall that $v_{\mathrm{rad},i}$ is the measured velocity of star $i$ ($i\in[1,N]$) and $\epsilon_{\mathrm{v_{rad}},i}$ its uncertainty. The likelihood $\mathcal{L}$ of the model parameters is the product of the individual probabilities. Following \citet{2002AJ....124.3270G}, we calculate the quantity $\lambda\equiv-2\ln\mathcal{L}$ and obtain
\begin{equation}
 \label{eq:lambda_stat}
 \lambda = \sum_{i=1}^N \ln \left[ 2\pi(\sigma_\mathrm{los}^2+\epsilon_{\mathrm{v_{rad}},i}^2) \right] + \sum_{i=1}^N \frac{(v_{\mathrm{rad,}i}-v_\mathrm{sys})^2}{\sigma_\mathrm{los}^2+\epsilon_{\mathrm{v_{rad}},i}^2}\,.
\end{equation}
It can be shown that $\lambda$ follows a $\chi^2$ distribution with $N$ degrees of freedom around the expectation value,
\begin{equation}
 \label{eq:lambda_expval}
 \langle\lambda\rangle = \sum_{i=1}^N \ln \left[ 2\pi(\sigma_\mathrm{los}^2+\epsilon_{\mathrm{v_{rad}},i}^2) \right] + N\,.
\end{equation}
Whether or not a model is statistically acceptable can thus be verified using the well-known characteristics of the $\chi_N^2$-distribution.

Furthermore, likelihood ratio tests can be applied to discriminate between individual models in a statistical manner. We assume that the likelihood of models has been calculated on an $m$-dimensional grid and that the model that maximizes the likelihood has $\lambda_\mathrm{min}$. If the range of models is restricted to an ${\hat m}$-dimensional subspace, the quantity $\lambda-\lambda_\mathrm{min}$ will be $\geq$0 for each model in ${\hat m}$. A well-known theorem from statistical theory \citep{1938AMS.....9...60W} states that for large sample sizes $N$, the difference $\lambda-\lambda_\mathrm{min}$ is distributed according to a $\chi^2$-distribution with $m-{\hat m}$ degrees of freedom. Therefore, confidence intervals can also be obtained using $\chi^2$ statistics. This approach was also followed, e.g., by \citet{1993ApJ...409...75M} or \citet{2002AJ....124.3270G}.

\bibliographystyle{aa}
\bibliography{pmas_clusters_merged_arxiv_v1}

\end{document}